\documentclass[11pt,a4paper]{article}
\usepackage{jheppub}
\usepackage{tikz}

\usepackage{graphicx,verbatim}
\usepackage{amsmath}
\usepackage{amsfonts}
\usepackage{amssymb}
\usepackage{psfrag}
\usepackage{accents}
\usepackage{mathrsfs}
\usepackage[colorinlistoftodos]{todonotes}
\usepackage[normalem]{ulem}
\usepackage{epstopdf}

\def\be{\begin{equation}}
\def\ee{\end{equation}}
\def\ba{\begin{eqnarray}}
\def\ea{\end{eqnarray}}

\def\={\hat{=}}

\def\f{\frac}

\def\Lie{\mathcal{L}}
\def\rmd{\mathrm{d}}
\newcommand{\pb}[1]{\hbox{\lower0.8ex\hbox{${}_{\leftarrow}$}}\kern-1.9ex{#1}}
\newcommand{\const}{\mathrm{const}}

\def\BMS{\mathfrak{B}}

\def\Spi{\mathfrak{S}}

\def\B{\mathcal{B}}

\def\qo{\mathring{q}}
\def\go{\mathring{g}}
\def\Co{\mathring{c}}
\def\Ko{\mathring{k}}

\def\Do{\mathring{D}}

\def\no{\mathring{n}}

\def\ello{\mathring{\ell}}
\def\sigmao{\mathring{\sigma}}

\def\omegao{\mathring{\omega}}

\def\kappao{\mathring{\kappa}}

\def\rmd{\mathrm{d}}

\newcommand*{\scri}{\ensuremath{\mathscr{I}}} 
\newcommand*{\scrip}{\ensuremath{\mathscr{I}^{+}}} 
\newcommand*{\scrim}{\ensuremath{\mathscr{I}^{-}}} 
\newcommand*{\scripm}{\ensuremath{\mathscr{I}^{\pm}}} 
\makeatletter
\newcommand*{\rom}[1]{\expandafter\@slowromancap\romannumeral #1@}
\makeatother

\def\inot{i^\circ}
\def\Poincare{\rm Poincar\'e\,\,}
\def\bsP{P_{\!a}^{\rm BS}}
\def\admP{P_{\!a}^{\rm ADM}}
\def\inotP{\mathfrak{p}_{\inot}}
\def\bmsP{\mathfrak{p}^{\rm bms}_{\inot}} 
\def\spiP{\mathfrak{p}^{\rm spi}_{\inot}} 
\def\scriJ{\vec{J}_{\!\scrip}} 
\def\inotJ{\vec{J}_{\inot}} 

\def\Mo{\mathring{M}}
\def\Omegao{\mathring{\Omega}}
\def\go{\mathring{g}}
\def\nablao{\mathring{\nabla}}
\def\no{\mathring{n}}
\def\lo{\mathring{\ell}}
\def\Co{\mathring{C}}
\def\Ko{\mathring{K}}
\def\uo{\mathring{u}}
\def\No{\mathring{N}}
\def\Vo{\mathring{V}}
\def\Vo{\mathring{V}}

\def\bfbeta{\boldsymbol{\beta}}
\def\bfepsilon{\boldsymbol{\epsilon}}

\def\bfzeta{\boldsymbol{\zeta}}
\def\bftau{\boldsymbol{\tau}}
\def\AM{\emph{AM}\,}
\def\v{\mathbf{v}}
\def\n{\mathbf{n}}

\begin{document}
\title{Unified Treatment of null and Spatial Infinity IV: Angular Momentum at Null and Spatial Infinity}
\author[a]{Abhay Ashtekar,}
\emailAdd{ashtekar.gravity@gmail.com}
\author[b]{Neev Khera}
\emailAdd{nkhera@uoguelph.ca}
\affiliation[a]{Institute for Gravitation and the Cosmos \& Physics Department, Penn State, University Park, PA 16802, U.S.A.\\
Perimeter Institute for Theoretical Physics, 31 Caroline St N, Waterloo, ON N2L 2Y5, Canada}

\affiliation[b]{Physics Department, University of Guelph, Guelph, Ontario, N1G 2W1, Canada}

\abstract{
\noindent
In a companion paper \cite{aank1} we introduced the notion of \emph{asymptotically Minkowski spacetimes}. These space-times are asymptotically flat at both null and spatial infinity, and furthermore there is a harmonious matching of limits of certain fields as one approaches $\inot$ in null and space-like directions. These matching conditions are quite weak but suffice to reduce the asymptotic symmetry group to a \Poincare group $\inotP$.
Restriction of $\inotP$ to future null infinity $\scrip$ yields the canonical \Poincare subgroup $\bmsP$ of the BMS group $\BMS$ selected in \cite{etnrp,aa-rad} and that its restriction to spatial infinity $\inot$, the canonical subgroup $\spiP$ of the Spi group $\Spi$ selected in \cite{aarh,aa-ein}. As a result, one can meaningfully compare angular momentum that has been defined at $\inot$ using $\spiP$ with that defined on $\scrip$ using $\bmsP$. We show that the angular momentum charge at $\inot$ equals the sum of the angular momentum charge at any 2-sphere cross-section $S$ of $\scrip$ and the total flux of angular momentum radiated across the portion of $\scrip$ to the past of $S$. In general the balance law holds only when angular momentum refers to ${\rm SO(3)}$ subgroups of the \Poincare group $\inotP$.
}

\maketitle

\section{Introduction}
\label{s1}

Isolated systems in general relativity with vanishing cosmological constant are modeled by asymptotically flat spacetimes. The nature of radiative aspects of the gravitational field becomes transparent as one recedes from sources in null directions \cite{bondi,sachs,sachs2,rp}, while the Coulombic aspects reveal themselves as one recedes in space-like directions \cite{adm,rg-jmp}. These simplifications have led to a detailed understanding of asymptotic symmetries and the associated conserved quantities in each regime. However, the two descriptions are disparate. A framework to unify them was introduced in \cite{aarh,aa-ein} through the notion of AEFANSI space-times that are Asymptotically Empty and Flat at Null and Spatial Infinity. The goal of this paper --and its preceding companion \cite{aank1}-- is to fill the gaps that still remained. 

Specifically, the AEFANSI framework provides a 4-dimensional description of spatial infinity without a 3+1 split and introduces $\scri$ as the light cone of $\inot$, thereby providing a strategy to unify the two descriptions. However, asymptotic conditions themselves were imposed only as $\inot$ is approached along space-like directions. In the companion paper \cite{aank1} we introduced additional requirements as one approaches $\inot$  in \emph{null} directions  --i.e., along $\scrip$\,-- by a natural continuation of the conditions that already hold along space-like directions. 
Space-times satisfying our strengthened asymptotic conditions are called Asymptotically Minkowskian (\AM). They provide a natural arena to discuss isolated gravitating systems --such as compact binaries that merge, or scatter off each other--  for which the space-time geometry is expected to be asymptotically flat in both regimes, not just separately but in a harmonious manner. 

Our strengthening of asymptotic conditions in the passage from AEFANSI to \AM space-times is geometrically natural, and also mild in the sense that it refers only to a small set of fields. At the same time, it is rather subtle in that it refers to the limit in which space-like \emph{directions} at $\inot$ approach null directions, avoiding the use of the hyperboloid of \emph{unit space-like vectors} at $\inot$ that is sometimes used for this purpose. As a result, restrictions on curvature as one approaches $\inot$ are weaker. For example, the previous works assumed that the Weyl components that enter the integrands of charges on $\scrip$ have finite limits as one approaches $\inot$, which are moreover continuous in the `infinite boost limit' in which space-like directions of approach become null \cite{Prabhu:2019fsp,kpis}. With our boundary conditions, they can well diverge in the infinite boost limit, as is expected in generic physically interesting situations. The energy momentum and angular momentum charges are still finite because our analysis involves limits of charge \emph{integrals} rather than their integrands and the divergent terms in the integrands integrate out to zero before the limit is taken.

In the distant past, isolated gravitating systems --such as binaries-- are generally well-described by the post-Newtonian framework in which there is a single \Poincare group at infinity that energy momentum and angular momentum of the system refer to. It is therefore natural to ask if this \Poincare group extends to the phase of evolution in which the binaries are sufficiently close, when a full general relativistic treatment is required. Only then can one meaningfully compare linear and angular momenta at early, intermediate and late times. 

Now, it has been known for some time that in AEFANSI space-times, one can extract a canonical \Poincare subgroup $\spiP$ of the Spi group $\Spi$ by imposing a rather weak restriction on the fall-off of the magnetic part of the Weyl tensor as one approaches $\inot$ along space-like directions. (Energy-momentum and angular momentum normally defined using initial data --e.g. in the post-Newtonian framework-- refer to this \Poincare group \cite{am3+1}.) Similarly,  one can extract a canonical \Poincare subgroup $\bmsP$ of the BMS group $\BMS$ by requiring an additional, rather weak condition on the fall-off of the magnetic part of the Weyl tensor, but now as one approaches $\inot$ along $\scrip$ \cite{etnrp,aa-rad}. One can use $\spiP$ to define energy momentum and angular momentum charges at $\inot$ and, similarly, $\bmsP$ to define them on any cross-section $S$ of $\scrip$. However, since $\spiP$ is a subgroup of $\Spi$ and $\bmsP$ a subgroup of $\B$, the two \Poincare groups have remained disparate. To meaningfully compare the two sets of charges, one needs a natural isomorphism between $\bmsP$ and $\spiP$. 

In general AEFANSI space-times, using the tangent space at $\inot$ one \emph{can} set up a natural isomorphism between the \emph{4-dimensional translation subgroups} of $\BMS$ and $\Spi$. This made it possible to analyze the relation between the ADM 4-momentum ${\admP}$ at $\inot$ and the Bondi-Sachs 4-momentum $\bsP[S]$ defined on any cross-section of $S$ of $\scrip$ \cite{aaam-prl}. On the other hand, angular momentum $\inotJ$ at $\inot$ refers to a ${\rm SO(3)}$ subgroup of $\spiP$, while angular momentum $\scriJ[S]$ evaluated at a cross-section $S$ of $\scrip$ refers to a ${\rm SO(3)}$ subgroup of $\BMS$, and there are \emph{infinitely many} ${\rm SO(3)}$ subgroups in $\BMS$ (any two being related by a combination of a supertranslation and a boost). Therefore, given the ${\rm SO(3)}$ subgroup of $\spiP$ that $\inotJ$ refers to, one has to \emph{first} isolate a canonical ${\rm SO(3)}$ subgroup of $\BMS$ that corresponds to the one used to define $\inotJ$, and \emph{then} analyze the relation between $\inotJ$ and $\scriJ[S]$ (associated with that canonical subgroup). Now, in stationary space-times one \emph{can} set up a natural isomorphism between the specific groups used in the two regimes and show that the two angular momenta are equal  \cite{aams-jmp}. (Recall that in stationary space-times there are no gravitational waves to carry away angular momentum.) The goal of this paper is to extend that analysis to general \AM space-times introduced in \cite{aank1} which admit generic gravitational waves. 

This will be possible because in any \AM space-time, each generator $\zeta^a$ of the asymptotic symmetry group $\inotP$ of \AM space-times is automatically a generator of $\bmsP$ on $\scrip$, and of $\spiP$ at $\inot$. Therefore, given the rotational generators $\tilde\zeta^a$ that are used to define $\inotJ$, one can unambiguously locate the corresponding generators $\zeta^a$ of $\bmsP$, define $\scriJ [S]$ using them, and investigate the relation between $\inotJ$ and $\scriJ$. We will find that the charge $\inotJ$ at $\inot$ is the sum of the charge $\scriJ[S]$ at the cross-section $S$ of $\scrip$ and the flux of $\vec{J}$ across the portion of $\scrip$ to the past of $S$. This is precisely the expected balance law. However, it holds \emph{only} for ${\rm SO(3)}$ generators that belong to $\inotP$. If for example we choose a vector field $\tilde\xi{}^a$ which is in the Lie-algebra of $\spiP$ at spatial infinity but \emph{not} in the Lie-algebra of $\bmsP$ on $\scrip$, the equality fails to hold because $\tilde\xi{}^a$ does not satisfy the regularity conditions that hold for generators of asymptotic symmetries of \AM space-times. Indeed, since  $\tilde\xi{}^a$  would not belong to $\inotP$, from the perspective of \AM space-times it is not even meaningful to compare the two charges; it would be like comparing apples with oranges.

The paper is organized as follows. In section \ref{s2} we begin by recalling the definitions of angular momentum charges at null and spatial infinity. It has been known for some time that the expression of charge integrals $\scriJ [S]$, evaluated at a cross-section $S$ of $\scrip$, can be obtained by first appropriately extending generators $\zeta^a$ of BMS rotations to vector fields $\tilde{\zeta}^a$ in a neighborhood of $\scrip$, then constructing the `Komar integrals' using these $\tilde{\zeta}^a$ on a family of 2-spheres $\hat{S}$ in the physical space-time, and finally taking the limit of these Komar integrals as the 2-spheres $\hat{S}$ smoothly converge to the given cross-section $S$ of $\scrip$ \cite{geroch1981linkages}. This is the well-known `linkage formalism' \cite{winicour1968some}. Using results of \cite{am3+1}, we extend it to include 2-spheres that converge to $\inot$ and show that in the limit Komar integrals now yield $\inotJ$. Section 3 contains the main result of this paper.  In section \ref{s3.1} we isolate the ${\rm SO(3)}$ subgroups of $\spiP$ and $\bmsP$ that are used to define $\inotJ$ and $\scriJ$ respectively. In section \ref{s3.2} we introduce a continuous family of cylinders $C$ on which the conformal factor $\Omega$ is constant and which are bounded by 2-spheres $\hat{S_1}$ and $\hat{S}_0$, where $\hat{S}_0$ lie on a partial Cauchy surface $\Sigma$ that passes through $\inot$ (see Fig. \ref{fig:foliation}). Now, by Stokes' theorem, the difference between the Komar integrals evaluated on $\hat{S}_0$ and $\hat{S}_1$ on any one $\Omega= {\rm const}$ cylinder $C$ can be expressed as a 3-surface `flux' integral across $C$. In the limit as $\Omega \to 0$, the 2-spheres $\hat{S}_1$ converge to a cross-section $S_1$ of $\scrip$ and the 2-spheres $\hat{S}_0$ converge to $\inot$. By construction, the extended charge integrals converge to  $\scriJ[S_1]$ and $\inotJ$ respectively.  In section \ref{s3.3} we show that the flux integral across cylinders $C$ converges to the flux of angular momentum across the portion of $\scrip$ to the past of $S$. The argument is rather subtle in that it involves an interchange of the operations of taking the limit $\Omega \to 0$ and performing the 3-surface integral over $C$, which is justified thanks to the \AM boundary conditions and the fact that the diffeomorphism generated by $\tilde{\zeta}^a$ is in the \Poincare group $\inotP$. In section \ref{s4} we summarize the results and put them in a broader perspective.

Our conventions are as follows. The physical metric is denoted by $\hat{g}_{ab}$, the conformal metric of the \AM completion by $g_{ab}$, and of completions in which $\scrip$ is divergence-free by $\go_{ab}$. We use  -,+,+,+ signature. The torsion-free derivative operator of $g_{ab}$ is denoted by $\nabla$ and curvature tensors are defined via:  $2\nabla_{[a}\nabla_{b]} k_c = R_{abc}{}^d k_d$, $R_{ac} = R_{abc}{}^b$, and $R=g^{ab}R_{ab}$. In case of an ambiguity, will use the equality \, $\=$\, to emphasize that the equality holds only at $\scrip$ or at $\inot$. Away from $\inot$, we simply assume that the fields are appropriately smooth. For example, $g_{ab}$ is assumed to be $C^4$ for simplicity, although as discussed in \cite{aank1}, lower differentiability at $\scrip$ suffices for much of the discussion. In the main part of the discussion, we assume that there is no matter field in a neighborhood of $\scrip\cup\inot$. As we discuss in section \ref{s3.3.4}, it is rather straightforward to weaken this assumption to  allow, e.g., matter fluxes across $\scrip$.

We will conclude this Introduction with a brief discussion of the asymptotic conditions we use. In AEFANSI space-times, as one approaches $\inot$ along space-like directions, the conformally rescaled metric $g_{ab}$ is continuous, and the limits of its connection $\nabla$ are smooth in their angular dependence, but have radial discontinuities (that capture the value of the ADM 4-momentum). This endows $g_{ab}$ with a $C^{>0}$ differential structure of \cite{aarh}. As mentioned above, to define angular momentum at $\inot$ one has to  strengthen these conditions slightly by requiring that the magnetic part of the Weyl curvature should decay faster than what is guaranteed by the $C^{>0}$ condition on the metric (i.e., that limits to $\inot$ of\,  ${}^{*}C_{abcd}\, \nabla^b\Omega^{\f{1}{2}}\, \nabla^d\Omega^{\f{1}{2}}$ should exist, again, along space-like approach to $\inot$). In the passage to \AM space-times, used in this paper, one imposes, in addition, certain continuity requirements on limits to $\inot$ as  the space-like directions are `infinitely boosted' and become tangential to $\scrip$. The precise conditions are spelled out in detail in \cite{aank1} and summarized in section \ref{s3.3.3}. Complementary aspects of the current status of the existence of solutions satisfying these conditions are discussed in \cite{aank1} and in Section \ref{s4}.

\section{Asymptotic charges as limits of 2-sphere integrals in space-time}
\label{s2}

This section is divided into two parts. In the first we recall the definitions of asymptotic charges defined on cross-sections $S$ of $\scrip$ and at $\inot$. While individual results are well-known, charges at $\scrip$ are generally expressed using conformal completions in which the null normal to $\scrip$ is divergence-free, and in the discussion of charges at $\inot$, using  conformal completions in which the divergence of its null normal is positive (so that $\scrip$ is the future light cone of $\inot$). We make the relation between the two explicit. In the second part we recall that the angular momentum charges can be recovered as limits of 2-sphere Komar integrals, constructed from a suitable extension of the rotational symmetry vector field from $\scrip$ to the space-time interior. 

\subsection{Angular momentum charges}
\label{s2.1}

We will use the same notation as in the companion paper \cite{aank1}. Thus, physical space-times will be denoted by $(\hat{M}, \hat{g}_{ab})$, and their conformal completions satisfying the asymptotic conditions in the definition of \AM space-times by $(M, g_{ab} = \Omega^2 \hat{g}_{ab})$. In these completions, the metric $g_{ab}$ is $C^{>0}$ at $\inot$ and smooth elsewhere on $M$, and $\scri$ is the light cone of $\inot$. On the other hand, as noted above, in the literature that discusses $\scrip$ alone, one generally uses conformal completions $(\Mo, \go_{ab} = \Omegao^2\hat{g}_{ab})$ that make $\scrip$ divergence-free, i.e. for which $\nablao_ a \nablao^a \Omegao\, \equiv\, \nablao_a \no^a \,\=\, 0$, where $\=$ stands for `equals at $\scrip$'. Einstein's equations satisfied by $\hat{g}_{ab}$ then imply that $\no^a$ is in fact covariantly constant at $\scrip$, i.e., $\nablao_a \no_b\, \= \,0$. The conformal metric $\go_{ab}$ provides a canonical degenerate metric $\qo_{ab}$ on $\scrip$ via pull-back, as well as a canonical null normal $\no^a = \go^{ab} \nablao_b \Omegao$ to $\scrip$.%
\footnote{For a discussion of the conformal rescalings $g_{ab} = \omegao^2\go_{ab}$ between the \AM and divergence-free conformal completions, see section 3.1 of \cite{aank1}. As in \cite{aank1}, we have set $\no_a = \nablao_a \Omega$ in a divergence-free conformal completion and $n_a = \nabla_a \Omega$ in the \AM completion. Thus, these 1-forms are null only at $\scrip$  where they can serve as a part of the Newman-Penrose null tetrad.}

Let us recall the definitions of BMS charges using divergence-free conformal completions. Given any cross-section $S$ of $\scrip$, one can decompose any BMS vector field $V^a$ on $\scrip$ into a part that is along the null normal $\no^a$ and a part that is tangential to $S$:
\begin{equation}
   V^a = \big(f + \uo \kappao\big) \no^a\, +\, \zeta^a\,,
\end{equation}
where $f$ is a function on $\scrip$ satisfying $\Lie_{\no} f = 0$,\, $\uo$ is an affine parameter of $\no^a$ (which can be taken to be constant on $S$), $\kappao$ is given by $\Lie_{\no} V^a = \kappao\, \no^a$, and $\zeta^a$ is tangential to $S$, satisfying $\Lie_{\no} \zeta^a =0$ on $\scrip$. Let $\ello_a$ denote the null normal to $S$ satisfying $\no^a\ello_a = -1$. Then, the shear of the cross-section is given by $\sigmao_{ab} = \mathring{\underbar{q}}_a{}^c \mathring{\underbar{q}}_b{}^d \nablao_a \ello_d$ where $\mathring{\underbar{q}}_a{}^c$ is the projector into $S$. The charge\, $Q_V[S]$ \,evaluated at the cross-section $S$ of $\scrip$ is a linear map\, $V^a \to  Q_{V}[S]$\, from the Lie algebra of the BMS group to real numbers, given by (see, e.g., \cite{dray,tdms,adlkJ}):
\begin{align}
    \label{scriQ} 
     Q_{V}[S] :=  -\f{1}{8\pi G}\,\oint_S \Big[& \big( f + \uo \kappao \big)\,\big(\Ko_{abcd}\ello^a \no^b \ello^c\no^d + \frac{1}{2}{\sigmao}_{ab} {\No}_{cd}\, \mathring{\underbar{q}}^{ac} \mathring{\underbar{q}}^{bd}\big) 
    \nonumber\\ 
    & \phantom{=}+ \Ko_{abcd} \ello^a \no^b \ello^c \zeta^d + \zeta^b {\sigmao}_{ab}\, \mathring{\underbar{D}}_c{\sigmao}^{ac}\,+\,\textstyle{\f{1}{4}} \zeta^c \mathring{\underbar{D}}_c({\sigmao}_{ab} {\sigmao}^{ab}) \Big] \,\, \rmd^2 \Vo \, .
\end{align}
Here $\Ko_{abcd}$ is the limit of $\Omegao^{-1}\Co_{abcd}$ to $\scrip$, with $\Co_{abcd}$ the Weyl tensor of $\go_{ab}$; $\No_{ab}$ is the News tensor; $\mathring{\underbar{q}}_{ab}$ the intrinsic metric on $S$; and  $\mathring{\underbar{D}}$, the  derivative operator on $S$ compatible with $\mathring{\underbar{q}}_{ab}$. 
These charges satisfy a balance law: given any two cross-sections $S_1$ and $S_2$ of $\scrip$, we have
\be \label{balance1} Q_{V}[S_1] - Q_{V}[S_2]\, = \,\f{1}{16\pi G} \int_{\scrip_{[1,2]}}\!\! \big[(\Lie_V \Do_a - \Do_a \Lie_{V}) \ello_b\, +\,  2 \ello_{(a} \Do_{b)} \kappao \big]\, \No_{cd}\,\qo^{ac} \qo^{bd}\,
\, \rmd\uo \, \rmd^2\Vo\,  \ee
where $\scrip_{[1,2]}$ is the portion of $\scrip$ bounded by $S_1$ and $S_2$;\, $\qo^{ab}$ is any inverse of $\qo_{ab}$, the pull-back to $\scrip$ of $\go_{ab}$;\, $\Do$ is the derivative operator induced on $\scrip$ via pull-back of the space-time $\nablao$ (that is compatible with $\go_{ab}$);\, and $\ello_a$ is any covector field on $\scrip$ satisfying $\no^a\ello_a =-1$ that is orthogonal to both $S_1$ and $S_2$. 
%
%

Angular momentum $\scriJ[S]$ refers to the charge associated to an ${\rm SO(3)}$ subgroup of $\BMS$. 
Let us first fix a ${\rm SO(3)}$ subgroup and choose a cross-section $S$ of $\scrip$ to which generators $\zeta^a$ of this ${\rm SO(3)}$ subgroup are tangential. For these BMS vector fields the angular momentum charges 
$J_\zeta [S]$ are given by the Dray-Streubel expression \cite{dray,tdms}
\begin{equation}
     \label{scriJ1} 
     J_\zeta [S]\, =  -\f{1}{8\pi G}\,\oint_S \big[\Ko_{abcd} \ello^a \no^b \ello^c \zeta^d + \zeta^b\sigmao_{ab}\, \mathring{\underbar{D}}_c\sigmao^{ac}\,+\,\textstyle{\f{1}{4}}\, \zeta^c \mathring{\underbar{D}}_c(\sigmao_{ab} \sigmao^{ab}) \big] \,\, \rmd^2 \Vo \,.
\end{equation}
Our differentiability assumptions on $\go_{ab}$ imply that the fields in the integrand of (\ref{scriJ1}) are all $C^1$ on $\scrip$ whence the integral is manifestly well-defined. We can also simplify the balance law (\ref{balance1}) for these ${\rm SO(3)}$ subgroups. First, since $\zeta^a$ are rotations, without loss of generality, we can choose a conformal completion such that $\kappao \,\= \,0$. Next, given any two cross-sections $S_1$ and $S_2$ to which $\zeta^a$ is tangential, $\Lie_\zeta \ello_a$ vanishes on these cross-sections. Therefore, we can always choose $\ello_a$ (satisfying $\ello_a \no^a =-1$) for which $\Lie_\zeta \ello_a =0$ on the entire portion $\scrip_{[1,2]}$ of $\scrip$ bounded by $S_1$ and $S_2$.  Then the balance law (\ref{balance1}) reduces to 
\be \label{balance2}  J_{\zeta}[S_1] - J_{\zeta}[S_2]\, = \, \f{1}{16\pi G} \int_{\scri^+_{[1,2]}} \big[\Lie_\zeta\sigmao_{ab}\, ]\,\No_{cd}\,\,  \mathring{\underbar{q}}^{ac} \mathring{\underbar{q}}^{bd}\, \rmd\uo \, \rmd^2\Vo\, , \ee
where $\mathring{\underbar{q}}^{ab}$ is now the inverse of $\qo_{ab}$ satisfying $\mathring{\underbar{q}}^{ab} \ello_b =0$. If, in addition, $\Lie_{\no} \lo_a =0$, then one can drop the last condition on the inverse metric and one has:
\be \label{balance3}  J_{\zeta}[S_1] - J_{\zeta}[S_2]\, = \, \f{1}{16\pi G} \int_{\scri^+_{[1,2]}} \big[\Lie_\zeta\sigmao_{ab}\, ]\,\No_{cd}\,\,  \qo^{ac} \qo^{bd}\, \rmd\uo \, \rmd^2\Vo\,  \ee 
where $\qo^{ac}$ is any inverse of $\qo_{ab}$.


As discussed above, to obtain the conformal completion that incorporates all the requirements of an \AM space-time, we need to make an appropriate conformal transformation $\go_{ab} \to g_{ab} = \omegao^2 \go_{ab}$, realizing $\scrip$ as the  light cone of the point $\inot$. Under this transformation, the expression of $J_\zeta [S]$ is recast to the form
\begin{equation}
     \label{scriJ2} 
     J_\zeta [S]\, =  -\f{1}{8\pi G}\,\oint_S \big[K_{abcd} \ell^a n^b \ell^c \zeta^d + \zeta^b\sigma_{ab}\, D_c\sigma^{ac}\,+\,\textstyle{\f{1}{4}}\, \zeta^c D_c(\sigma_{ab} \sigma^{ab}) \big] \,\, \rmd^2 V \,
\end{equation}
where $K_{abcd} = \lim_{\rightarrow \scrip} \Omega^{-1} C_{abcd} =\omegao^{-1}\Ko_{abcd}$,\, $n^a\, =\, \nabla^a \Omega\, \= \,\omegao^{-1} \no^a$, \, $\ell^a \,\= \,\omegao^{-1} \ello^a$, \,  $D$ is the derivative operator compatible with the intrinsic metric on the 2-sphere $S$, $\sigma_{ab}$ is the shear of $\ell_b$ on $S$. 
Thus, the form of the expression of $J_\zeta [S]$ remains unchanged in the passage from the conformal completion gives by $\go_{ab}$ to that given by $g_{ab}$ (thanks to the specific form of the last two terms). \\

\emph{Remark}: There are other angular momentum expressions discussed in the literature that also provide a linear map from the BMS generators $\zeta^a$ to real numbers but differ from (\ref{scriJ1}) in the numerical coefficients in front of the last two terms (for a summary, see \cite{Compere:2019gft,compere2021classical}). However, they suffer from two drawbacks. First, in general axisymmetric space-times with non-vanishing $N_{ab}$ they lead to non-zero angular momentum fluxes around the symmetry axis, which is physically unacceptable. Second, they do not admit a local flux, whence their angular momentum charge does not change continuously with continuous deformations of the cross-section $S$, which is mathematically awkward \cite{Chen:2022fbu}. The Dray-Streubel charge (\ref{scriJ1}) is free from of these limitations: In axisymmetric space-times yields the conserved Komar integral, and it has the continuity property because it was in fact obtained by integrating a \emph{local} flux \cite{aams}. There is also another candidate for angular momentum charges in the literature \cite{chen2021supertranslation}, obtained from taking limits to $\scrip$ of quasi-local angular momentum integrals in space-time. It also satisfies the continuity property, but in axisymmetric space-times with $N_{ab}\neq0$, there is a non-zero flux of angular momentum between generic cross-sections \cite{Chen:2022fbu}. Also, the quasi-local charge is not a linear mapping from the BMS Lie algebra to reals. Therefore, we will work with (\ref{scriJ1}) and (\ref{scriJ2}). \\ 
 
Let us now recall the definition of angular momentum at spatial infinity. In this case, the \AM boundary conditions ensure that the magnetic part of the Weyl tensor falls-off appropriately for one to single out a \Poincare subgroup $\inotP$ of the BMS group. Angular momentum charges $\inotJ$ are associated with ${\rm SO(3)}$ subgroups of this $\inotP$ (see section \ref{s2.2} for further details). Generators of rotations are represented by  vector fields $\bfzeta^a$  on the hyperboloid $\mathcal{H}$ of unit space-time directions at $\inot$. They preserve the intrinsic metric on $\mathcal{H}$ and, on any cross-section $S_0$ to which a given $\bfzeta^a$ is tangential, it has the form $\bfzeta^a = \bfepsilon^{ab}\, \mathbf{D}_b f$, where $\bfepsilon_{ab}$ is the area 2-form on $S_0$. The angular momentum charge associated with $\zeta^a$ is given by:
\be\label{Jinot1} J_\zeta[\inot] =   -\f{1}{8\pi G} \oint_{S_0} \bfbeta_{ac}\, f \,\bftau^a \bftau^c\, \rmd^2 \Vo \, ,\ee
where $\bfbeta_{ac} =  \lim_{\to \inot} \, C_{abc}{}^d\, \nabla^a \Omega\, \nabla_d \Omega$ \, is the leading, non-trivial contribution to the magnetic part of the Weyl tensor as $\inot$ is approached in space-like directions,\,  $\bftau^a$ is the unit normal within $\mathcal{H}$ to $S_0$, and $\rmd^2 \Vo$ is the volume element on a unit 2-sphere. This expression can be recast to a more familiar form in terms of fields that refer to the physical metric $\hat{g}_{ab}$, which will be more directly useful for our purposes. Let us begin by considering a partial Cauchy surface $\Sigma$ passing through $\inot$ which is $C^{>1}$ at $\inot$ and smooth elsewhere (as in Fig. \ref{fig:foliation}), foliated by 2-spheres $\hat{S}_0$ on which $\Omega = {\rm const}$. Denote by $\tilde{\zeta}^a$ vector fields in a neighborhood of $\inot$ that induce the given rotational symmetry $\bfzeta^a$ on $\mathcal{H}$. Without loss of generality we can assume that $\tilde{\zeta}^a$ are tangential to the 2-spheres $\hat{S}_0$. 
As $\Omega \to 0$, this family of 2-spheres converges to $\inot$ and the expression (\ref{Jinot1}) can be expressed as a limit of integrals on $\hat{S}_0$ involving the Cauchy data on $\Sigma$, induced by the \emph{physical} metric $\hat{g}_{ab}$ \cite{am3+1}:
\be \label{Jinot2} J_\zeta [\inot] = -\f{1}{8\pi G} \, \lim_{\Omega\to 0}\, \oint_{\hat{S}_0} \hat{K}_{ab}\, \tilde\zeta^a\, \hat\eta^b \rmd^2 \hat{V} \ee
where $\hat{K}_{ab}$ is the extrinsic curvature of $\hat\Sigma$ w.r.t. the physical metric $\hat{g}_{ab}$, \, $\tilde\zeta^a$ is an asymptotic symmetry vector field near $\inot$, tangential to the $\Omega={\rm const}$ 2-spheres,  that induces the desired rotation $\zeta^a$ in the Lie algebra of $\inotP$, and $\hat\eta^a$ is the unit normal to the family $\hat{S}$ of 2-spheres. We will use (\ref{Jinot2}) to relate $\inotJ$ to $\scriJ [S]$.

This concludes our summary of angular momentum charges at $\scrip$ and $\inot$. In the next subsection, using the `linkage framework' \cite{geroch1981linkages,winicour1968some} we will use a convenient extension of the symmetry vector fields $\zeta^a$ in a neighborhood of $\scrip$ and show that the expression of angular momentum (\ref{scriJ2}) at $\scrip$ and  (\ref{Jinot2}) at $\inot$ can both be expressed as the limits of the \emph{same} 2-sphere integrals in space-time. 

\subsection{Extensions}
\label{s2.2}

Let us extend the rotational BMS vector field $\zeta^a$ on $\scrip$ smoothly to a vector field $\tilde{\zeta}^a$ in a neighborhood $\mathfrak{N}$ of $\scrip\cup \inot$ in $M$ such that: (i) $\Lie_{\tilde\zeta} \Omegao = 0$,  and,\,  (ii) the Geroch-Winicour condition \cite{geroch1981linkages}
\begin{equation}
    \label{eq:div-free}
    \hat{\nabla}_a \tilde{\zeta}^a = 0\,
\end{equation}
is satisfied, in $\mathfrak{N}$. (Condition (i) implies that $\tilde{\zeta}^a$ is divergence-free also w.r.t. $\go_{ab}$.) Consider a smooth family of 2-spheres $\hat{S}$ that converge to a cross-section $S$ of $\scrip$ to which $\tilde{\zeta}^a$ is everywhere tangential. Then, the `linkage' charge $L_{\tilde\zeta} [S]$ is given by \cite{geroch1981linkages,winicour1968some}:
\begin{equation}
    \label{eq:linkage}
    \hat{L}_{\tilde\zeta}[\hat{S}] = \frac{1}{16\pi G}\oint_{\hat{S}} \hat{\epsilon}_{ab}{}^{mn} \,\hat{\nabla}_m \hat{\zeta}_n  \,\rmd S^{ab}\, .
\end{equation}
Here and in what follows, $\hat\zeta_a = \hat{g}_{ab}\,\tilde\zeta^b$; \emph{the hat in $\hat\zeta_b$ emphasizes that the index is lowered using the physical metric $\hat{g}_{ab}$}.\, $\hat{L}_{\tilde\zeta}[\hat{S}]$ has two interesting properties:
\begin{enumerate}
    \item The limit of $L_{\tilde\zeta}[\hat{S}]$ as $\hat{S}\to S$ is well-defined (even though $\hat{\zeta}_a$ diverges in the limit to $\scrip$); and,
    \item When $\zeta^a$ is tangential to $S$, the limit agrees with $J_{\tilde\zeta} [S]$ of Eq. \eqref{scriJ1} in any divergence-free conformal completion $(\Mo,\, \go_{ab})$ (and hence also with \eqref{scriJ2} in an \AM conformal completion $(M,\, g_{ab})$). 
\end{enumerate} 
Note that the integral on the right side is precisely the Komar integral defined by $\tilde{\zeta}^a$ in the physical space-time $(\hat{M}, \hat{g}_{ab})$. \\

\emph{Remark}: While linkages do not lead to physically acceptable charges and fluxes for BMS supertranslations, as noted above, with our overall normalization they do yield the correct angular momentum $J_\zeta [S]$. The linkage expressions used in \cite{geroch1981linkages} omitted overall numerical factors. 
The numerical factors in \cite{10.1063/1.525283} were tailored to give the standard 4-momentum while the ones used in this paper are tailored to give the standard angular momentum. Thus, the charge of Eq. (\ref{eq:linkage}) is half that used in~\cite{10.1063/1.525283}. \\ 

Let us now turn to $\inot$. While linkages have been analyzed extensively in the literature on null infinity, this is not the case for their limit to $\inot$. Let us therefore begin by recalling the salient features of the structure at $\inot$ in  AEFANSI space-times \cite{aarh,aa-ein}. Note first that metrics $g_{ab}$ and $\bar{g}_{ab}$ that provide conformal completions of an AEFANSI space-time $(\hat{M}, \hat{g}_{ab})$ are related by $\bar{g}_{ab}\, =\, \omega^2 g_{ab}$,\, where $\omega=1$ at $\inot$ and is $C^{>0}$ at $\inot$ and smooth elsewhere. Let us now restrict ourselves to those AEFANSI space-times in which the magnetic part of the asymptotic Weyl curvature falls-off faster than the electric part (i.e., as $1/r^4$ rather than $1/r^3$ in the physical space-time, a condition that is automatically satisfied in \AM space-times). Then, one can naturally select a preferred class of conformal completions in which the \emph{relative conformal factor} $\omega$ relating any two metrics is $C^{1}$ at $\inot$, (although each metric itself is only $C^{>0}$ there). The subset of vector fields $\tilde{\zeta}^a$ in the physical space-times representing infinitesimal Spi symmetries that \emph{in addition} leave this preferred class $\{g_{ab}\}$ of conformal completions invariant is severely restricted: these $\tilde{\zeta}^a$ generate only a \Poincare subgroup $\inotP$ of the infinite dimensional Spi group $\Spi$. Angular momentum $\inotJ$  refers to ${\rm SO(3)}$ subgroups of  $\inotP$. 

Fix one of these rotational subgroups and, as in section \ref{s2.1}, consider a partial Cauchy surface $\Sigma$ passing through $\inot$ which is $C^{>1}$ at $\inot$ and smooth elsewhere. It is naturally foliated by 2-spheres $\hat{S}_0$ on which $\Omega= {\rm const}$. Without loss of generality we can assume that the extension $\tilde{\zeta}^a$ of any given rotation generator of the fixed ${\rm SO(3)}$ subgroup is tangential to this foliation. In the discussion of $\inotJ$ this condition replaces the two assumptions we made on the extension in the discussion of linkages (although, we can also require those two conditions without loss of generality).  Let us perform a 3+1 decomposition of the linkage charge (\ref{eq:linkage}) using vector fields $\hat{\tau}^a$ and $\hat{\eta}^a$, where $\hat{\tau}^a$ is the (time-like) unit normal to $\Sigma$ and $\hat{\eta}^a$, the (space-like) unit normal to the 2-sphere foliation within $\Sigma$. Then we have:
\begin{align}
    L_\zeta[\hat{S}] &= -\frac{1}{16\pi G} \oint_{\hat{S}}2\hat{\tau}^{[a} \hat{\eta}^{b]}\, \hat{\nabla}_a \hat{\zeta}_b  \,\rmd^2 \hat{V} \,,\\
    &=  \frac{1}{16\pi G} \oint_{\hat{S}} \left( \hat{\tau}^a\tilde{\zeta}^b\,\hat{\nabla}_a \hat{\eta}_b -\hat{\eta}^{a}\tilde{\zeta}^{b}\,\hat{\nabla}_a \hat{\tau}_b\right)  \,\rmd^2 \hat{V}\,.
\end{align}
Now, since $\hat{\eta}_a$ is normal to the foliation by $\hat{S}_0$ 2-spheres, the pullback of $\hat\nabla_{[a} \hat\eta_{b]}$  to the 2-spheres $\hat{S}_0$ vanishes. Therefore, we have $\hat{\tau}^a\tilde{\zeta}^b\hat{\nabla}_a \hat{\eta}_b = \hat{\tau}^a\tilde{\zeta}^b\hat{\nabla}_b \hat{\eta}_a$. This allows us to rewrite the linkage charge as
\begin{align}
    L_{\tilde\zeta}[\hat{S}_0] &=  -\frac{1}{8\pi G} \oint_{\hat{S}_0}  \hat{\eta}^{a}\tilde{\zeta}^{b}\,\hat{\nabla}_{(a} \hat{\tau}_{b)}  \,\rmd^2 \hat{V} \,\\
    &=  -\frac{1}{8\pi G} \oint_{\hat{S}_0}  \hat{\eta}^{a}\tilde{\zeta}^{b}\, \hat{K}_{ab}  \,\rmd^2 \hat{V} \,,\label{eq:linkage_ang_3+1} 
\end{align}
where $\hat{K}_{ab}$ is the extrinsic curvature of the slice $\Sigma$. The limit of the right side of ~\eqref{eq:linkage_ang_3+1} as $\hat{S}_0$ converges on $\inot$ is exactly the 3+1 form (\ref{Jinot2}) of the angular momentum $J_\zeta [\inot]$ at spatial infinity. 

To summarize, given an appropriate extension $\tilde{\zeta}^a$ of asymptotic rotational symmetries $\zeta^a$ to a neighborhood of $\scrip \cup \inot$, limits of linkage (or, Komar) charges evaluated on a suitable family of 2-spheres $\hat{S}$ in the physical space-time yield the angular momentum charge $\scriJ [S]$ in the limit as $\hat{S}$ tend to the cross-section $S$ of $\scrip$, and $\inotJ$ in the limit they tend to $\inot$ along the partial Cauchy slice $\Sigma$.

\section{Flux and the balance law}
\label{s3}

In this section we will restrict ourselves to \AM space-times and relate $\scriJ [S]$ and $\inotJ$ using the fact that they can both be obtained as limits of linkage charges. The section is divided into three parts. In the first we specify the ${\rm SO(3)}$ subgroup of $\inotP$ that will be used to define angular momentum charges at $\scrip$ and $\inot$. 
In the second, we discuss the flux integrals on suitable portions of $\Omega = {\rm const}$ 3-manifolds that result from applying Stokes' theorem to the linkage charge integrals, and introduce specific structures in a neighborhood of $\scrip\cup \inot$ that will facilitate the task of taking the limits $\Omega\to 0$. In the third, we will take the limit of charge and flux integrals and establish the angular momentum balance law. 
%
\subsection{Selection of the preferred rotation generators}
\label{s3.1}

Fix an \AM space-time $(\hat{M},\, \hat{g}_{ab})$ and consider its conformal completions $(M, g_{ab})$ satisfying the boundary conditions at both $\inot$ and $\scrip$, introduced in \cite{aank1}. As pointed out in section~\ref{s1}, the asymptotic symmetry group of \AM space-times is the \Poincare group $\inotP$ whose restriction to spatial infinity yields the group $\spiP$, and whose restriction to $\scrip$ yields $\bmsP$. 

Let us begin with $\inot$ and summarize a convenient characterization of the generators of $\spiP$. 
Since $\inot$ is a single point, to discuss the defining properties of these generators we need to consider vector field $\tilde\zeta^a$ in a neighborhood of $\inot$. Let $\mathfrak{N}$ denote such a neighborhood. Recall from section~\ref{s2.1} that in \AM space-times, $\mathfrak{N}$ is equipped with a preferred equivalence class $\{g_{ab}\}$ of conformally related metrics whose \emph{relative} conformal factor $\omega$ is $C^1$ at $\inot$ (although each metric is only $C^{>0}$ there). The $\tilde\zeta^a$ in the Lie algebra of $\spiP$ generate diffeomorphisms that leave $\inot$ and the metric $g_{ab}$ at $\inot$ invariant (and thus induce a Lorentz rotation in the tangent space $T_{\inot}$ of $\inot$), and preserve the class $\{g_{ab}\}$. These requirements translate to certain conditions on the symmetry vector fields $\tilde\zeta^a$. They have to be $C^{1}$ at $\inot$, smooth elsewhere, and satisfy the following equations at $\inot$:

\be \label{zeta}
 (i)\,\,\,\, \tilde\zeta^a\, \=\,0;\qquad (ii)\,\,\,\, \nabla_{(a} \tilde\zeta_{b)}\, \=\,0; \qquad {\rm and},\qquad (iii)\,\,\,\, \nabla_c \big(\nabla_{(a} \tilde\zeta_{b)}\big)\, \= \,2\,(\nabla_c\phi)\, g_{ab} 
\ee 
where $\phi$ is a function on $\mathfrak{N}$ that is $C^1$ at $\inot$ and smooth elsewhere and, as usual, $\tilde\zeta_b = g_{ab} \tilde{\zeta}^a$. The three conditions follow directly from the three conditions that the diffeomorphsims have to satisfy to belong to $\spiP$. For details, see \cite{aa-ein,aank1}. 

In section~\ref{s2}, we defined angular momentum $\inotJ$ using any ${\rm SO(3)}$ subgroup of $\spiP$ and showed that it could be obtained as a limit of linkage/Komar integrals on an appropriate family of 2-spheres $\hat{S}_0$ in the physical space-time that converge to $\inot$. To relate $\inotJ$ with $\scriJ[S]$, we need to match the ${\rm SO(3)}$ subgroups in the two regimes. The idea is to use the asymptotic rest frame of the system to select these subgroups. 

Recall first that ADM 4-momentum $\admP$ is well-defined in \AM space-times and defines a time-like vector at $\inot$ \cite{10.1063/1.531497,10.4310/jdg/1669998184} (we assume that matter sources in space-time interior, if any, satisfy the necessary energy conditions). Consider the ${\rm SO(3)}$ subgroups of $\spiP$ that leave $\admP$ invariant. They are the rotation subgroups defined by the asymptotic rest-frame at $\inot$  and are thus related by the action of space-translations in $\spiP$. The corresponding generators $\tilde{\zeta}^a$ that induce the same rotation in $T_\inot$ but differ in the values of $\nabla_c \big(\nabla_{(a} \tilde\zeta_{b)}\big)$ at $\inot$, i.e., lead to different vectors $\nabla_a\phi$ of Eq. (\ref{zeta}) at $\inot$. Because the ADM 3-momentum vanishes in this asymptotic rest frame, these ${\rm SO(3)}$ subgroups define the same $\inotJ$. Nonetheless, given a conformal completion, it is convenient to remove the ambiguity in the choice of the ${\rm SO(3)}$ subgroup \emph{by requiring that $\nabla_a\phi$ should vanish at $\inot$}. In the asymptotic rest frame of the given completion, we can think of these preferred $\tilde\zeta^a$ as `pure rotations', uncontaminated by translations (since translations $\hat{t}^a$ in the Lie algebra of $\spiP$ are characterized by $\nabla_a \phi$ at $\inot$). 


Next, let us turn to null infinity. Now, in \AM space-times the magnetic part ${}^{\star}\!{K}_{ab}$ of the asymptotic Weyl curvature goes to zero also as one approaches $\inot$ along $\scrip$ \cite{aank1}. This property in turn ensures that $\scrip$ admits a 3-parameter family of foliations by 2-sphere cross-sections $S$ with following properties: (i) shear of $S$ goes to zero in the asymptotic past in any foliation; (ii) any two leaves of one of these preferred foliations are related by a BMS \emph{time}-translation; (iii) the group of BMS translations acts simply and transitively on the 4-parameter family of cross-sections; and, (iv) a generic supertranslation (i.e. one that is \emph{not} a translation) maps any one of these cross-sections out of the 4-parameter family. These properties hold both in \AM conformal completions $(M, g_{ab})$ and divergence-free conformal completions $(\Mo, \go_{ab})$. The requirement that asymptotic symmetries should preserve this 4-parameter family of preferred cross-sections (in addition to the universal structure of $\scrip$) weeds out supertranslations and reduces $\BMS$ to the \Poincare group $\bmsP$ \cite{etnrp,aank1}. 

In our discussion of linkages in section~\ref{s2} there is complete freedom in the choice of ${\rm SO(3)}$ subgroups also at null infinity: we could have chosen \emph{any} ${\rm SO(3)}$ subgroup of $\BMS$ and defined the corresponding angular momentum $\scriJ[S]$ via Eq. (\ref{scriJ1}). However, as discussed above, to relate $\scriJ$ to $\inotJ$, we need to restrict this choice. A natural avenue is to use the past limit of the Bondi-Sachs 4-momentum $\bsP [S]$ evaluated on any leaf $S$ of our preferred foliations: The desired ${\rm SO(3)}$ subgroups of $\bmsP$ are those whose action on $\scrip$ leaves this past limit invariant. 

To make the action of these subgroups on $\scrip$ explicit, let us use divergence-free conformal completions as in the literature that discusses $\scrip$ by itself. Any of these conformal completions $(\Mo, \go_{ab})$ provides a pair of fields $(\qo_{ab}, \no^a)$ intrinsically defined on $\scrip$, where $\qo_{ab}$ is the (degenerate) metric on $\scrip$ and\, $\no^a\, \= \,\go^{ab}\nablao_b \Omegao$\, is a null normal to $\scrip$. These pairs will be referred as \emph{conformal frames}. In a general conformal frame, while $\no^a$ is a `time-like' super-translation, it need not be BMS \emph{translation}. Those in which it is a time translation are called Bondi conformal frames (and the corresponding $\qo_{ab}$ turns out to be a round 2-sphere metric). There is a unique Bondi conformal frame in which the past-limit of $\bsP [S]$ is purely  time-like, i.e., in which the spatial 3-momentum $\vec{P}^{\rm BS} [S]$ vanishes in the infinite past. Consider any of the preferred foliations that is left invariant by the time translation $\no^a$ of this Bondi conformal frame. The ${\rm SO(3)}$ subgroups of $\bmsP$ of interest are those for which the generators $\zeta^a$ are tangential to the leaves of any one of these foliations. Again, there is a space-translation ambiguity in the choice of these preferred foliations --and hence of the preferred ${\rm SO(3)}$ subgroups-- defined by $\no^a$, but the past limit of $\scriJ$ is insensitive to this freedom.

In this discussion, we used divergence-free conformal completions only to cast the procedure in a more familiar language. The entire construction --the preferred foliations and the use of the past limit of $\bsP [S]$-- can be readily restated using \AM completions in which the null normal $n^a$ has positive divergence so that $\inot$ is a point, and $\scrip$ its future light cone. To eliminate the remaining freedom in the choice of the ${\rm SO(3)}$ subgroup of $\bmsP$, let us pass to an \AM completion. Then, the preferred ${\rm SO(3)}$ subgroups leave the ADM 4-momentum $\admP$ invariant, since the past limit of $\bsP [S]$ along $\scrip$ agrees with $\admP$ \cite{aaam-prl}. The linkage extensions $\tilde\zeta^a$ of generators $\zeta^a$ of the preferred ${\rm SO(3)}$ subgroups automatically satisfy the first two conditions in (\ref{zeta}) as one approaches $\inot$ along $\scrip$. However, in general the right side of the third condition is not zero. The ${\rm SO(3)}$ subgroup of $\bmsP$ that matches the unique subgroup we selected to define $\inotJ$ is the one whose generators $\zeta^a$ admit an extension $\tilde\zeta^a$ for which the right side vanishes, i.e.,  
\be \label{consistency}
\nabla_c \big(\nabla_{(a} \tilde\zeta_{b)}\big)\, \= \,0 \qquad {\rm at} \,\, \inot \ee
not only as one approaches it along space-like directions, but also along $\scrip$. (See the remark at the end of this subsection.) This requirement ensures that the rotation generated by $\tilde\zeta^a$ in $\bmsP$ is the same as that it generates in $\spiP$. Put differently, this matching condition at $\inot$ ensures that $\tilde\zeta^a$ does not just define an element of the Lie algebra of $\bmsP$ on $\scrip$ and, separately,  an  element of the Lie algebra of $\spiP$; it ensures that the rotations in the two groups are appropriately matched so that $\tilde\zeta^a$ yields a well-defined  element of the Lie algebra of $\inotP$. Since angular momenta $J_\zeta [S]$ at $\scrip$ and $J_\zeta[\inot]$ at $\inot$ will be defined using these rotation generators (in (\ref{scriJ2}) and (\ref{Jinot2}) respectively), the two angular momenta can be meaningfully compared. Finally, note that this ${\rm SO(3)}$ subgroup of $\bmsP$ picks out a \emph{unique} foliation of $\scrip$, each leaf of which is left invariant by this ${\rm SO(3)}$ subgroup. (Moreover, the leaves $S$ become asymptotically shear-free and are mapped to each other by the BMS translation that is aligned with the past limit of $\bsP [S]$).\\

\emph{Remark}: The geometrical underpinning that leads to (\ref{consistency}) can be summarized as follows. The metric $g_{ab}$ at $\inot$ in any permissible conformal completion of an \AM space-time is universal and the conformal factor relating them is $C^{1}$ at $\inot$. Hence, if $g_{ab}$ and $\bar{g}_{ab} = \omega^2 g_{ab}$ are two allowed completions, $\omega \=1$ and $C^{1}$ at $\inot$. The two metrics can be distinguished by the limits to $\inot$ of their derivative operators $\bar\nabla$ and $\nabla$ and the difference $(\bar{\nabla}_a - \nabla_a)$ is encoded in the convector $\nabla_a\omega$ evaluated at $\inot$. (For further details, see \cite{aa-ein,aank1}.) If the 1-parameter family of diffeomorphisms generated by $\tilde\zeta^a$ (and parameterized by $\lambda$) lies in the \Poincare group $\inotP$, then the metric $g_{ab}$ is sent to $\bar{g}_{ab}(\lambda) = \omega^2(\lambda) g_{ab}$, so that under the infinitesimal motion generated by $\tilde\zeta^a$ we have $\Lie_{\tilde{\zeta}} g_{ab}\, \= \,2\phi g_{ab}$ at $\inot$ where $\phi  = \f{d\omega(\lambda)}{d\lambda}\mid_{\lambda=0}$. If  $\tilde\zeta^a$ belongs to the Lie-algebra of our unique ${\rm SO(3)}$ subgroup in the given conformal completion $g_{ab}$, then $\phi \=0$\, and\, $\omega(\lambda) =1$. That is, under the action of this rotation subgroup, the derivative operator $\nabla$ is left invariant in the limit as one approaches $\inot$ along space-like directions. Now, in \AM space-times the derivative operator $\nabla$ is continuous: One obtains the same limit whether one first goes to $\scrip$ and then takes the limit to $\inot$, or if one first approaches $\inot$ along space-like directions and then boosts the direction to become null. Therefore, the fact the 1-parameter family of diffeomorphisms leaves the limit of $\nabla$ along space-like direction unchanged implies that $\nabla$ is also left invariant as $\inot$ is approached along $\scrip$: $\lim_{\to \inot}\, (\Lie_{\tilde{\zeta}} \nabla_a - \nabla_a \Lie_{\tilde{\zeta}}) k_b\mid_{\scrip}\,=0$ for all 1-forms $k_b$. Expanding this equation and using the fact that $k_b$ is arbitrary, one obtains $\lim_{\to \inot}\, \big(\nabla_a \nabla_b \tilde{\zeta}_c = \tilde\zeta^m R_{mabc}\big)\!\mid_{\scrip}$. Finally, symmetrization on $b$ and $c$ implies  (\ref{consistency}). 

To summarize, because $\tilde\zeta^a$ generates a rotation in $\inotP$, and elements of $\inotP$ preserve all the asymptotic conditions satisfied by the \AM space-time $(M, g_{ab})$,\, $\tilde \zeta^a$ satisfies (\ref{zeta}) not only along space-like approach to $\inot$ but also null.

\subsection{Flux associated with the linkage charge}
\label{s3.2}

\begin{figure}
    \centering
    \includegraphics{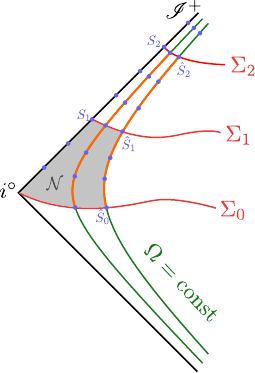}
    \caption{Extension of structures in $(M, g_{ab})$ from $\scrip \cup \inot$ to the space-time interior. $\scrip$ is foliated by a preferred family of 2-sphere cross-sections $S$ that become shear-free in the asymptotic past. A 1-parameter family of space-like 3-manifolds $\Sigma$ intersect $\scrip$ at these 2-spheres $S$.\, $\Sigma_0$ is the limiting 3-manifold of the $\Sigma$-family that passes through $\inot$ and is $C^{>1}$ there.  The $\Omega={\rm const}$ (time-like) 3-manifolds intersect the space-like 3-manifolds $\Sigma$ in 2-spheres $\hat{S}$ that converge to the cross-sections $S$ of $\scrip$, as $\Omega \to 0$. Along $\Sigma_0$, the 2-spheres $\hat{S}_0$ converge to $\inot$. The (gray) shaded region denotes the closed neighborhood $\mathcal{N}$ of $\inot$ enclosed by the surfaces $\Sigma_0$, $\Sigma_1$, $\scrip$ and the $\Omega=\const$ surface that passes through $\hat{S}_0$ and $\hat{S}_1$.}
\label{fig:foliation}
\end{figure}

As we saw in section~\ref{s2.2}, the linkage charge $L_\zeta [\hat{S}]$ is a 2-sphere integral in space-time that yields both $J_\zeta[S]$ at null infinity and $J_\zeta [\inot]$ at spatial infinity in appropriate limits. We now wish to choose the family of 2-spheres $\hat{S}$ in a convenient manner to make the relation between the two angular momenta transparent. Let us fix an \AM completion $(M, g_{ab})$ of the given physical space-time. Let $S$ denote leaves of the unique foliation of $\scrip$ picked out at the end of section~\ref{s3.1}. The ${\rm SO(3)}$ subgroup of interest leaves each leaf of this foliation invariant.

Let us foliate a neighborhood $\mathfrak{N}$ of $\scrip \cup \inot$ by $\Omega= {\rm const}$ (time-like) 3-manifolds. Let us also introduce a partial Cauchy surface $\Sigma$ in $\mathfrak{N}$ passing through $\inot$, that is orthogonal to $\admP$ and $C^{>1}$ at $\inot$ and smooth elsewhere. Denote the 2-spheres at which the $\Omega= {\rm const}$ surfaces intersect $\Sigma$ by $\hat{S}_0$. Next, let us introduce another smooth foliation of the portion of $\mathfrak{N}$ to the future of $\Sigma$ by a family of space-like surfaces that intersects $\scrip$ in the leaves $S$ of the preferred foliation, with $\Sigma$ serving as the past boundary leaf. The 2-sphere intersections of these space-like surfaces with the $\Omega= {\rm const}$ surfaces will be denoted by $\hat{S}$. (See Fig. \ref{fig:foliation}.) Generators $\zeta^a$ of the preferred ${\rm SO(3)}$ subgroup of $\inotP$ are, by construction, tangential to the leaves $S$ of the preferred foliation of $\scrip$. Let us extend them as vector fields $\tilde{\zeta}^a$ to the portion of $\mathfrak{N}$ of interest, subject to the following conditions:\, (i) the extension satisfies the Geroch-Winicour condition $\hat\nabla_a  \tilde\zeta^a =0$;\, (ii) $\tilde\zeta^a$ are tangential to the 2-spheres $\hat{S}$ (including the $\hat{S}_0$ that lie on $\Sigma$). In particular, then, $\Lie_{\tilde\zeta} \Omega =0$, whence $\nabla_a \tilde{\zeta}^a =0$ as well.  

With this structure at hand, let us consider the linkages,  $L_{\tilde{\zeta}} [\hat{S}_1]$ and $L_{\tilde{\zeta}} [\hat{S}_0]$ associated with a general 2-sphere $\hat{S}_1$ in our family, and $\hat{S}_0$, the 2-sphere that lies on $\Sigma$, both on the same $\Omega = {\rm const}$ surface. As we saw in section~\ref{s2.2}, in the limit $\Omega \to 0$ (and moving along the leaves of the space-like foliation), $\hat{S}_1$ tends to a preferred cross-section $S_1$ of $\scrip$ and $\hat{S}_0$ to $\inot$. In this limit,  $L_{\tilde{\zeta}} [\hat{S}_1]$ yields the angular momentum $J_\zeta [S]$ at $\scrip$, and  $L_{\tilde{\zeta}} [\hat{S}_0]$, the angular momentum $J_\zeta[\inot]$. Therefore, it is instructive to first express $L_{\tilde{\zeta}} [\hat{S}_0]$ as the sum of  $L_{\tilde{\zeta}} [\hat{S}_0]$ and a 3-surface flux integral across the cylindrical portion  $C$ of the 
$\Omega = {\rm const}$ surface, bounded by $\hat{S}_1$ and $\hat{S}_0$, and then investigate the limit of the flux integral.

Let us then take the exterior derivative of the charge aspect in~\eqref{eq:linkage} and integrate it along $C$. We have 
\be \label{linkagebalance} L_{\tilde{\zeta}}[\hat{S}_1] - L_{\tilde{\zeta}}[\hat{S}_0] = F_{\tilde{\zeta}}[C]\, ,\ee 
where the flux  $F_{\tilde{\zeta}}[C]$ across $C$ is given by
\begin{align}
    F_{\tilde{\zeta}}[C] &= \frac{3}{16\pi G}\int_C \hat{\epsilon}_{[ab}{}^{cd}\hat{\nabla}_{e]}\, \hat{\nabla}_c\,\hat{\zeta}_d \,\, \rmd S^{abe}\\
    &=-\,\frac{1}{32\pi G}\int_C \hat{\eta}^{f}\,\hat{\epsilon}_{f}{}^{abe}\,\,\hat{\epsilon}_{ab}{}^{cd}\,\hat{\nabla}_{e} \hat{\nabla}_c\,\hat{\zeta}_d \,\, \rmd^3\hat{V} \,. \\
\end{align}
Here, in the second line we have replaced $\rmd S^{abe}$\, by\, $\f{1}{3!}\,\hat\epsilon^{abe}\,\rmd^3\hat{V}$  and, as before, set $\hat\zeta_a = \hat{g}_{ab} \tilde\zeta^b$ and denoted the unit radial normal to the $\Omega={\rm const}$ surface $C$ by $\hat\eta^a$. Next, by expanding out the  contraction of the two volume forms and using the identity
\begin{equation}
    \hat{\nabla}_{e} \hat{\nabla}_c\hat{{\zeta}}_d = \hat{R}_{dce}{}^{k}   \hat{{\zeta}}_k + \hat{\nabla}_c \hat{\nabla}_{(e}\hat{{\zeta}}_{d)} +\hat{\nabla}_e \hat{\nabla}_{(c} \hat{{\zeta}}_{d)} - \hat{\nabla}_{d}\hat{\nabla}_{(e}\hat{{\zeta}}_{c)} \,,
\end{equation}
we can rewrite $F_{\hat{\xi}}[C]$ as 
\begin{align} \label{flux1}
    F_{\tilde{\zeta}}[C] &= -\frac{1}{8\pi G}\int_C  \hat{\eta}^d\,\hat{\nabla^e} \hat{\nabla}_{(e}\hat{{\zeta}}_{d)} \,\rmd^3\hat{V}\,,
\end{align}
where, we have used our assumption that there is no matter in the neighborhood of $\scri\cup\inot$ and thus set $\hat{R}_{ab}=0$. 

Next, let us recast the expression of $F_{\tilde{\zeta}}[C]$ using the conformally rescaled metric $g_{ab} = \omega^2 \hat{g}_{ab}$ which is well-behaved at $\scrip \cup \inot$.  Since  $\Lie_{\tilde{\zeta}} \Omega =0$ in the region under consideration,  the integrand on the right side of (\ref{flux1}) can be readily rewritten as: 
\be { \hat{\eta}^d\,\hat{\nabla^e} \hat{\nabla}_{(e}\hat{{\zeta}}_{d)} = \Omega\, (n\cdot n)^{-\f{1}{2}}\,\, n^d \big(\nabla^e \nabla_{(e} \tilde\zeta_{d)}\, -\, 4\, n^e\nabla_{(e} \tilde\zeta_{b)} \big)}\ee
where, as before, $n_a = \nabla_a\Omega$ and $\tilde\zeta_a = g_{ab} \tilde\zeta^b$. In the limit as $\Omega \to 0$, the cylinder $C$ tends to the portion $\scrip [S_1,\, \inot]$ of $\scrip$ to the past of $S_1$. Therefore, one would expect that the volume element $\rmd^3 \hat{V}$ would be simply related to $\rmd^3 V$ on $\scrip$. This expectation is correct, but there is a subtlety in defining $\rmd^3 V$ since $\scrip$ is null. Let us therefore
spell out the argument. The volume form on the time-like cylinder $C$ is\, $\rmd^3 \hat{V}= \hat{\epsilon}_{abc}\, \rmd^3 S^{abc}\, =\, - \hat\epsilon_{mabc}\, \hat\eta^m \,\rmd^3 S^{abc}$, while the standard volume 3-form $\epsilon_{abc}$ on $\scrip$ is given by $\epsilon_{mabc} = -4 n_{[m} \epsilon_{abc]}$. Therefore, we have: 
\ba \label{vol} 
\rmd^3 \hat{V} &=& \,  - \Omega^{-3} (n\cdot n)^{-\f{1}{2}}\, \epsilon_{mabc}\, n^m\, \rmd^3 S^{abc}\,=\,
4 \Omega^{-3} (n.n)^{-\f{1}{2}}\, n_{[m} \epsilon_{abc]}\, n^m\, \rmd^3 S^{abc}\nonumber\\
&=&\,\,\,\, \Omega^{-3}\, (n.n)^{\f{1}{2}}\,  \rmd^3 V \ea
Therefore, in terms of the conformally rescaled metric $g_{ab}$ we have:
\be \label{flux_aspect}\hat{\eta}^d\,\hat{\nabla^e} \hat{\nabla}_{(e}\hat{{\zeta}}_{d)} \,\rmd^3\hat{V}\, =\, \Omega^{-2} n^d\left(\nabla^e \nabla_{(e}\tilde{\zeta}_{d)}\,- \, 4\,\Omega^{-1}n^e \nabla_{(e}\tilde{\zeta}_{d)} \right) \rmd^3 V\, , \ee
and (\ref{flux1}) becomes:
\begin{align}
    \label{eq:flux-conformal}
    F_{\tilde{\zeta}}[C] &= -\frac{1}{8\pi G}\int_C  \Omega^{-2} n^d\left(\nabla^e \nabla_{(e}\tilde{\zeta}_{d)}\,- \, 4\,\Omega^{-1}n^e \nabla_{(e}\tilde{\zeta}_{d)} \right) \rmd^3 V \\
     &\equiv  \frac{1}{16\pi G}\,\int_C \mathcal{F}_{\tilde{\zeta}}\;\rmd^3 V\, ,
\end{align}
where the integrand of~\eqref{eq:flux-conformal} is the flux-aspect $\mathcal{F}_{\tilde{\zeta}}$.

Interestingly, the two terms that feature in $\mathcal{F}_\zeta [C]$ have a pleasing geometrical interpretation that will be used in section~\ref{s3.3}. Consider the 1-parameter family of diffeomorphisms generated by $\tilde\zeta^a$. The infinitesimal change $g^\prime_{ab}$ in the metric is of course given by $g^\prime_{ab} = \Lie_{\tilde\zeta} g_{ab} = 2\nabla_{(a} \tilde\zeta_{b)}$ (which vanishes both on $\scrip$ and at $\inot$). The connection $\nabla$ also changes under this diffeomorphism and the infinitesimal change $\nabla^\prime$ is given by 
\be \nabla_a^\prime k_b = C_{ab}^c\, k_c = \f{1}{2}\, g^{cd} \big[\nabla_a (\Lie_{\tilde\zeta}\, g_{bd}) +  \nabla_b (\Lie_{\tilde\zeta}\, g_{ad}) - \nabla_d (\Lie_{\tilde\zeta}\, g_{ab})\big]\, k_c \, ,\ee
which vanishes as we approach $\inot$ along space-like or null directions because our rotation generators preserve the connection $\nabla$ at $\inot$. Setting $k_c = \nabla_c \Omega \equiv n_c$,\,contracting with $g^{ab}$,\, and noting that the extension $\tilde\zeta^a$ of the symmetry vector field $\zeta^a$ on $\scrip$ satisfies $\nabla_a \tilde\zeta^a =0$ in the neighborhood $\mathcal{N}$, we obtain
\be g^{ab}\, \nabla_a^\prime n_b\, =\, n^b\, \nabla^a (\nabla_{(a}\tilde\zeta_{b)}) \qquad {\rm in}\,\,\,\, \mathcal{N}\, .  \ee
Therefore, the linkage flux aspect can be rewritten as an expression that involves only $\Omega$, the metric $g_{ab}$ and the infinitesimal changes $g^\prime_{ab}$ and $\nabla^\prime$ in the metric and the connection induced by the infinitesimal diffeomorphism generated by $\tilde\zeta^a$:
\be \label{flux2} \mathcal{F}_{\tilde\zeta} [C] = 2 \Omega^{-2}\, \big( g^{ab}\,\nabla_a^\prime n_b\, - 2 n^e\,n^d\, g^\prime_{ed}\big)\qquad {\rm in}\,\,\,\, \mathcal{N}\, . \ee
This geometrical recasting of $\mathcal{F}_{\tilde\zeta} [C]$ will be useful in section~\ref{s3.3} to take the limit 
$\Omega\to 0$.\\

\emph{Remark:}  Had we passed to a divergence-free conformal completion $(\Mo, \go_{ab})$ in place of $(M, g_{ab})$, 
the same argument that led us to Eq. (\ref{flux_aspect}) would have implied that 
\be \hat{\eta}^d\,\hat{\nabla^e} \hat{\nabla}_{(e}\hat{{\zeta}}_{d)} \,\rmd^3\hat{V}\, =\, \Omegao^{-2} \no^d\left(\nablao^e \nablao_{(e}{\mathring{\tilde{\zeta}}}_{d)}\,- \, 4\,\Omegao^{-1}\no^e \nablao_{(e}\mathring{\tilde\zeta}_{d)} \right) \rmd^3 \mathring{V}\, \equiv\, { 2}\, \mathring{\mathcal{F}}_{\tilde\zeta}. \ee
Geroch and Winicour~\cite{geroch1981linkages} showed that the flux $\mathring{\mathcal{F}}_{\tilde\zeta}$ in the divergence-free frame admits a smooth limit to $\scrip$. Since $\rmd^3 V = \omegao^3 \rmd^3 \Vo$ where the conformal factor $\omegao$ relating $\go_{ab}$ to $g_{ab}$ is smooth on $\scrip$, it follows that $\mathcal{F}_{\tilde\zeta}$ also admits a smooth limit to $\scrip$. We will use this fact in section \ref{s3.3}.

\subsection{Asymptotic limit of the balance law}
\label{s3.3}

The linkage balance law (\ref{linkagebalance}) ensures that  difference between $L_{\tilde\zeta}[\hat{S}_1]$ and $L_{\tilde\zeta}[\hat{S}_0]$ is the flux $\mathcal{F}_\zeta [C]$ across the cylinder $C$ joining $\hat[S_1]$ and $\hat{S}_0$. We already know that in the limit $\Omega \to 0$, the linkage charge $L_{\tilde\zeta}[\hat{S}]$ yields angular momentum charge $J_\zeta[S]$ at the cross-section $S$ of $\scrip$ and $L_{\tilde\zeta}[\hat{S}_0]$ provides the angular momentum charge $J_{\tilde\zeta} [\inot]$ at spatial infinity. The remaining task then is to take the limit of the flux integral.

\subsubsection{Statement of the problem and a strategy to address it}
\label{s3.3.1}

Let us proceed in two steps. First, let us consider only the portion $C_{{}_{[1,2]}}$ of the cylinder bounded by the cross-sections $\hat{S}_1$ and $\hat{S}_2$ that converges to the portion $\scri^+_{[1,2]}$ of $\scrip$ bounded by $S_1$ and $S_2$. Then, as we noted above, the linkage charges converge to the Dray-Streubel charges \cite{dray,tdms}  which satisfy the simplified version (\ref{balance3}) of the balance law at $\scrip$ because $\zeta^a, \, \no^a$ and $\ello_a$ meet all the conditions that led to the simplification. Therefore, the linkage flux integral (\ref{eq:flux-conformal}) over $C_{{}_{[1,2]}}$ converges to the flux across $\scri^+_{{}_{[1,2]}}$ provided by (\ref{balance3}). This balance law is expressed using a divergence-free conformal completion. Let us first re-express it using the \AM conformal completion using the fact that the \AM metric $g_{ab}$ is related to the metric $\go_{ab}$ in the divergence-free completion via $g_{ab} = \omegao^2 \go_{ab}$. For the fields at $\scrip$ that feature in the flux expression of Eq.(\ref{balance3}) we have 
\be   \sigma_{ab}\, \=\, \omegao \sigmao_{ab}; \qquad {q}^{ab} = \omegao^{-2} \qo^{ab}; \qquad \rmd^3 V = \omegao^3 \rmd \Vo\, . \ee
Since $\No_{ab}$ has zero conformal weight among divergence-free frames, let us set $N_{ab} := \No_{ab}$.  
Then, (\ref{balance3}) yields the following angular momentum balance law in the \AM conformal completion:
\be \label{eq:J_flux} J_\zeta[S_1] -  J_\zeta[S_2] = \f{1}{16\pi G}\, \int_{{\scri}^+_{[1,2]}}\!\! (\Lie_\zeta \sigma_{ab})\, N_{cd}\, {q}^{ac} {q}^{bd}\, \rmd^3 V\, , \ee
where $\scri^+_{[1,2]}$ is the closed portion of $\scrip$ with boundaries $S_1$ and $S_2$. 
To obtain the desired difference $J_{\tilde\zeta} [\inot] - J_\zeta[S_2]$, it is tempting to take the limit as $S_1$ recedes to infinite past and conclude that $J_{\tilde\zeta} [\inot] - J_\zeta[S_2]$ is given by replacing $\scri^+_{[1,2]}$ in the flux on the right-hand side by the portion  $\Delta\scrip$ between $\inot$ and $S_2$.
However, if one lets $S_1$ in (\ref{eq:J_flux}) to recede to infinite past, i.e, to $\inot$, one would obtain a balance law that is again \emph{intrinsic to $\scrip$}: It would simply say that difference between $J_\zeta[S_2]$ and the past limit of $J_\zeta[S_1]$ is given by the flux across $\Delta\scrip$. This procedure does not yield a relation between $J_\zeta[S_2]$ and $J_{\tilde\zeta} [\inot]$ because we do not know that $J_\zeta[S_1]$ will tend to $J_{\tilde\zeta} [\inot]$ as $S_1 \to \inot$!

The second step is to arrive at a strategy that does relate  $J_\zeta[S_2]$ to $J_{\tilde\zeta} [\inot]$. 
Let us return to the linkage balance law (\ref{linkagebalance}) and apply it to the cylinder $C$ between $\hat{S}_0$ and $\hat{S}_1$ and \emph{then} take the limit $\Omega\to 0$. Then we have
\be \label{prefinal} J_\zeta[S_1] -  J_{\tilde{\zeta}} [\inot]\,  = \, \lim_{\Omega \to 0}\, \int_{C} \mathcal{F}_{\tilde\zeta}\, \rmd^3 V \, .\ee
In the limit $\Omega \to 0$, $C$ converges to $\Delta\scri$ and, as remarked at the end of section \ref{s3.2}, the linkage flux $\mathcal{F}_{\tilde\zeta}$ extends smoothly to $\Delta\scri$ \cite{geroch1981linkages}. Since  $\inot$ is simply a point of zero measure, it can be removed from the volume integral. Thus, \emph{if we could interchange the limit and the integral}, we would be able to express the total flux across  $F_{\tilde{\zeta}}$ relating $J_\zeta[S_1]$ and  $J_{\tilde{\zeta}} [\inot]$ as an integral of $(\Lie_\zeta \sigma_{ab})\,N_{cd}\,{\underbar{q}^{ac}} {\underbar{q}^{bd}} $ over $\Delta\scrip$. This interchange would be justified if, for example, we were to replace $\inot$ with a cross-section $S_1$ as in the beginning of this section,  because $F_{\tilde{\zeta}}$ is smooth in the relevant space-time region and this region is compact. Then (\ref{eq:J_flux}) would be recovered as the limit of the linkage balance law as the cylinder $C$ bounded by $\hat{S}_2$ and $\hat{S}_1$ tends to the portion ${\scri}^+_{[1,2]}$ of $\scrip$ bounded by $S_2$ and $S_1$.

However, for the infinite cylinders $C$ bounded by $\hat{S}_2$ and $\hat{S}_0$, where the 2-spheres $\hat{S}_0$ coverage to $\inot$, the interchange of the limits and the integral is quite non-trivial. It can be justified when the flux integrand $\mathcal{F}_\zeta\, \rmd^3V$ is \emph{bounded} above by an integrable density in the closed neighborhood $\mathcal{N}$ of $\inot$, enclosed by the surfaces $\scrip$, $\Sigma_0$, $\Sigma_2$, and the $\Omega=\const$ surface passing through $\hat{S}_0$ and $\hat{S}_2$ (See Fig.~\ref{fig:foliation}). The difficulty is that $\mathcal{F}_{\tilde\zeta}\, \rmd^3 V$ \emph{fails to be smooth} $\inot$ whence, a priori, it may not be bounded in $\mathcal{N}$. We will establish the required boundedness in two steps. We will first show that, thanks to the conditions satisfied by the symmetry generators $\tilde\zeta^a$ of \AM space-times, $\mathcal{F}_{\tilde\zeta}\,\rmd^3V$ admits finite limits along every spatial and null direction approaching $\inot$. However, a priori, the limits along space-like directions may diverge as the limit in which they are infinitely boosted to null directions, causing an infinite discontinuity. In the second step, we will show that this does not happen; the integrand is bounded in $\mathcal{N}$, allowing us to interchange the integral and the limit and establish the desired balance law.

\subsubsection{Limits to $\inot$ of $\mathcal{F}_\zeta\, \rmd^3V$}
\label{s3.3.2}

Consider space-like curves $c(\lambda)$ passing through $\inot$, that is $C^{>1}$ there and smooth elsewhere, and take limits of various fields along them. Recall from section \ref{s3.1} that  $\tilde\zeta^a \, \=\,0$;\, $\nabla_{(a} \tilde\zeta_{b)}\, \=\,0$;\, and\, $\nabla_c (\nabla_{(a} \tilde\zeta_{b)})\, \=\,0$ at $\inot$. Therefore, fields $\Omega^{-1}\, (\nabla_{(a} \tilde\zeta_{b)})$ and $\Omega^{-\f{1}{2}}\, \nabla_c (\nabla_{(a} \tilde\zeta_{b)})$ admit regular direction dependent limits to $\inot$ \cite{aarh}. Similarly, limit of $\Omega^{-\f{1}{2}}\, \nabla^a \Omega = \Omega^{-\f{1}{2}}\,n^a$ is well-defined (and equals the tangent vector to $c(\lambda)$ at $\inot$ with norm $2$). Finally, because the 2-spheres of cylinders $C$ shrink to a point at $\inot$, \, $\Omega^{-1} \rmd^3 V$ as a well-defined limit. Therefore, it follows immediately that the linkage flux integrand $\mathcal{F}_{\tilde\zeta}\, \rmd^3 V$ has a well-defined  limit to $\inot$ along all space-like $c(\lambda)$ under consideration. 

Next, let us consider the limit of the flux aspect as we approach to $\inot$ along $\scrip$. Recall from the Remark at the end of section \ref{s3.2} that $\mathcal{F}_{\tilde\zeta}$ admits a smooth limit $\mathring{\mathcal{F}}_{\tilde\zeta}$ to $\scri$ in any divergence-free conformal completion (and hence also in any \AM conformal completion). Furthermore, as shown in \cite{10.1063/1.525283}, $\mathring{\mathcal{F}}_{\tilde\zeta}$ is related to the Hamiltonian flux on $\scrip$,
\be \mathring{\mathcal{H}}_{\tilde\zeta}\, :\= \,\f{1}{16\pi G}\, \big(\No^{ab}\,\Lie_\zeta \sigmao_{ab}\big)\mid_{{}_{\scrip}}\, ,\ee
also evaluated in a divergence-free completion, via  
\be \Lie_{\no} \big(\mathring{\mathcal{F}}_{\tilde\zeta} - \mathring{\mathcal{H}}_{\tilde\zeta}\big)\,\, \=\,\, \f{1}{64\pi G}\,\, \Lie_\zeta \big(\No^{ab} \No_{ab}\big)\, . \ee
Now, in an \AM space-time the news tensor is guaranteed to fall off as $1/\uo^2$ and shear $\sigmao_{ab}$ of our preferred slices as $1/\uo$ (in a divergence-free conformal completion) \cite{aank1}. Therefore, we conclude that in the $\uo \to -\infty$ limit along $\scrip$ the linkage flux $\mathring{\mathcal{F}}_{\tilde\zeta}$ has the asymptotic form $\mathring{f}(\theta,\varphi) + O(1/\uo)^3$ for some $\mathring{f}$ that has conformal weight $-3$. Now, $\rmd^3\Vo = \rmd \uo\, \rmd^2\Vo$ where $\rmd^2\Vo$ is the volume element of a unit 2-sphere. Therefore, $\mathring{\mathcal{F}}_{\tilde\zeta}\, \rmd^3\Vo$ has a well-defined limit as $\uo \to -\infty$. Since this combination is invariant under conformal rescalings, in an \AM completion ${\mathcal{F}}_{\tilde\zeta}\, \rmd^3 V$ also admits a well-defined direction dependent limit to $\inot$.\\ 

\emph{Remark}: As an aside, note that even if $\mathring{f}(\theta,\varphi) =0$, the linkage and Hamiltonian flux aspects differ on $\scrip$, although they would both tend to zero as $\uo \to -\infty$. Therefore, it may first appear that there is a contradiction with the equality in Eq.~(\ref{eq:J_flux}) that holds at $\scrip$, where by the left side equals the linkage flux, by definition, while the right side equals the Hamiltonian flux across $\scrip_{[1,2]}$. However, there is no contradiction because the form of the difference is such that it integrates out to zero on our 2-spheres because $\zeta^a$ is tangential to them. On the other hand, there \emph{would be} a contradiction if $\mathring{f}(\theta,\varphi)$ is not only non-zero but also has a constant part, i.e. if $\oint \mathring{f} \rmd^2 \Vo \not= 0$. Thus, while our argument does not show that the limits to $\inot$ of linkage and Hamiltonian flux aspects are equal, it implies that the difference between their 2-sphere integrals is constrained to vanish in the limit to $\inot$. 

\subsubsection{Boundedness of $\mathcal{F}_\zeta\, \rmd^3V$ and the desired balance law}
\label{s3.3.3}

We have shown that the integrand $\mathcal{F}_{\zeta}\, \rmd^3 V$ of the linkage flux admits a finite limit along any space-like or null direction approaching $\inot$. However, as  noted above, it is possible that the limit along a 1-parameter family of boosted space-like directions may diverge as they approach a null direction (creating an infinite discontinuity).  If this were to happen then we would not be able to interchange the integral and the limit. We will now show that, because the diffeomorphisms generated by $\tilde\zeta^a$ preserve the AM boundary conditions, the limit along the 1-parameter family of boosted space-like directions matches continuously with the limit along the null direction they approach, whence there is no divergence. 
For a general \AM space-time, this interchange would not be possible if $\tilde\zeta^a$ is \emph{not} a generator of the asymptotic symmetry group $\inotP$ because then the required boundedness will fail. 

Let us begin by recalling from \cite{aank1} the continuity conditions that ensure appropriate gluing of structures on $\scrip$ and at $\inot$ in \AM space-times. First, the metric $g_{ab}$ is continuous at $\inot$ in both space-like and null directions of approach. Second, the derivative operator $\nabla$ admits \emph{direction dependent} limits both in space-like and null directions. Next, consider any 1-parameter family of space-like curves $c_{\v}(\lambda)$ passing through $\inot$ that are $C^{>1}$ there with tangent vectors ${\v}^a$, smooth elsewhere, jointly continuous in $\v$ and $\lambda$, and converges to a curve $c_{\n}$ with a null tangent vector $\n$ as $\lambda \to 1$. Then, the limits $\nabla^{(\v)}$ to $\inot$ along $c_{\v}(\lambda)$ continuously approach the limit of $\nabla^{(\n)}$ along $c_{\n}$, as $\lambda \to 1$. The third continuity condition refers to the set $\mathcal{T}$ of tensor fields that (i) are constructed from the conformal factor $\Omega$, the metric $g_{ab}$ its derivative  $\nabla$ and fields constructed from them (e.g. curvature tensors and their derivatives); (ii) admit regular direction-dependent limits along space-like directions as one approaches $\inot$; and (iii) are $C^{1}$ in a neighborhood of $\scrip$ alone (not including $\inot$). (Thus, if two tensor fields belong to $\mathcal{T}$ so does their tensor product and contraction.) In \AM space-times, elements of $\mathcal{T}$ admit well-defined direction-dependent limits also along the limiting curves $c_{\n}$ of families $c_{\v}(\lambda)$ and, furthermore, the continuity property holds: The limits to $\inot$ along $c_{\v}(\lambda)$ converge to the limit along $c_{\n}$ as $\lambda \to 1$.

Now, since asymptotic symmetries preserve all boundary conditions, action of $\inotP$ must preserve this set. In particular, then, tensor fields representing the infinitesimal changes $g^\prime_{ab}$ and $\nabla^\prime$ induced by any of our rotation generators $\tilde\zeta^a$ belong to $\mathcal{T}$. Therefore, $ \mathfrak{F}_{\tilde\zeta} := \big( g^{ab}\,\nabla_a^\prime n_b\, - 2 n^e\,n^d\, g^\prime_{ed}\big)$ belongs to $\mathcal{T}$. Next, as we see explicitly from Eqs. (\ref{eq:flux-conformal}) and (\ref{flux2}), the linkage flux integrand is given by $\mathcal{F}_{\tilde\zeta}\, \rmd^3 V \, =\, 2\, \Omega^{-2}\,\mathfrak{F}_{\tilde\zeta}\, \rmd^3 V$. It would belong to $\mathcal{T}$ if and only if\, (i) it admits a regular direction dependent limit to $\inot$ along families of space-like curves $c_{\v}(\lambda)$ and (ii) is $C^1$ in a neighborhood of $\scrip$ alone. We established the first condition (i) in section~{\ref{s3.3.2}, and we know from \cite{geroch1981linkages} that the second condition (ii) holds at $\scrip$. Therefore, it follows that properties guaranteed by the continuity condition hold. The first property is that $\mathcal{F}_{\tilde\zeta}\, \rmd^3 V$ admit direction-dependent limits to $\inot$ along null directions. Indeed, we explicitly showed that this property holds (showing consistency with the present general argument). The second property is that, for any family of  curves $c_{v}(\lambda)$ mentioned above, the limits $\mathcal{F}_{\tilde\zeta}\, \rmd^3 V\,(\lambda)$ tend to the limit of $\mathcal{F}_{\tilde\zeta}\, \rmd^3 V$\, along curves $c_{\n}$. This implies that the limit does not diverge as  $\lambda \to 1$,  whence $\mathcal{F}_{\tilde\zeta}\, \rmd^3 V$ is bounded in $\mathcal{N}$ (including $\inot$).

Consequently, the interchange of the limit and integration in Eq. (\ref{prefinal}) is justified, and we have:
 \ba \label{balance4} J_\zeta[\inot] -  J_{\tilde{\zeta}} [S_2]\,  &=& \, \lim_{\Omega \to 0}\, \int_{C} \mathcal{F}_{\tilde\zeta}\, \rmd^3 V \, =\, \int_{\Delta\scrip}\, \lim_{\Omega \to 0}\,\big( \mathcal{F}_{\tilde\zeta}\,\,  \rmd^3 V\big)\, \nonumber\\
&=& \f{1}{16\pi G}\, \int_{\Delta\scrip}\,(\Lie_\zeta \sigma_{ab})\,N_{cd}\,{{q}^{ac}} {{q}^{bd}} \,\rmd^3 V \, . \ea
Here $\Delta\scrip$ is the portion of $\scrip$ to the past of $S_2$. Thus, we have the desired balance law:
The difference between the angular momentum $J_\zeta[\inot]$ is the sum of the angular momentum $J_{\tilde{\zeta}} [S_2]$ at a cross-section $S_2$ of $\scrip$ and the angular momentum flux across $\Delta\scrip$. Note that $J_\zeta[\inot]$ involves structures that refer \emph{only to} asymptotic flatness at spatial infinity (see Eqs. (\ref{Jinot1}) and (\ref{Jinot2})). Similarly, $J_\zeta[S_2]$ and the flux across $\Delta\scrip$ refer \emph{only to} asymptotic flatness at null infinity (see Eqs. (\ref{scriJ2}) and (\ref{balance2})). Individual terms in this balance law have no knowledge of asymptotic  flatness in the other regime.  Finally, since in the limit in which $S_2$ recedes to infinite past, the flux integral on the right side of (\ref{balance4}) vanishes, we conclude 
\be \label{past-limit1} 
\lim_{{S_2\to \inot}}\,\,J_{\tilde{\zeta}} [S_2] = J_\zeta[\inot] \ee
even though, again, fields entering the integrand on the two sides have no knowledge of other regime. The equality holds because in \AM space-times $\scrip$ and $\inot$ are glued appropriately.

The gluing conditions are geometrically motivated, and they are also quite weak. In particular, there is no assumption that the fields that enter the integrand of  $J_{\tilde{\zeta}}[S_2]$ of (\ref{scriJ2}) tend to those that enter the integrand of (\ref{Jinot1}) in the limit in which $S_2$ recedes to $\inot$. Indeed, this may not even hold! Furthermore, the integrand of (\ref{scriJ2}) (including the volume element $\rmd^2 V$)  generically diverges in this limit in scattering situations! This point becomes transparent in Bondi conformal frames in which the volume element $\rmd^2 \Vo$ remains unchanged as $S_2$ recedes. In these conformal completions, the component of the Weyl curvature that enters the integrand of (\ref{scriJ2}) --\,$\Psi_1^\circ$, in the Newman-Penrose notation--will diverge as $\uo \to -\infty$ in generic scattering situations. But this divergence is such that the limit of the 2-sphere integral $J_{\tilde{\zeta}}[S_2]$ is well-defined. More explicitly, the divergences can occur only in the $\ell>1$ components of $\Psi_1^\circ$ in the spin weighted spherical harmonic decomposition, while the integrand of (\ref{scriJ2}) is sensitive only to the $\ell=1$ components. And the equality $J_\zeta[\inot] = \lim_{S_2 \to \inot} J_{\tilde{\zeta}} [S_2]$ refers \emph{only to} the past limit of the \emph{integral} on the right side of (\ref{scriJ2}) and to the \emph{integral} on the right side of (\ref{Jinot1}) (or (\ref{Jinot2})). The \AM boundary conditions allowed us to establish the desired relation between the \emph{integrals} without any assumption on the relation between the two integrands. The strategy we used was to first extend the angular momentum charge integrals to the space-time interior appropriately, and then take the limit $\Omega\to 0$ of \emph{integrals} in the balance law on cylinders $\Omega = {\rm const}$ in the physical space-time.

\subsubsection{Remarks}
\label{s3.3.4}

We will now comment on various aspects of the main result.

1. In arriving at the balance law (\ref{balance4}), we extended structures at $\scrip\cup \inot$ to the space-time interior as shown in Fig.~\ref{fig:foliation}. The construction guaranteed that $S_2$ belongs to the family of preferred cross-sections that are shear-free in the infinite past, and, the rotation generator $\zeta^a$ is tangential to it. We will now show that the balance law continues to hold if we replace $S_2$ by \emph{any} smooth cross-section $S$ even when it violates both these restrictions. 

We will use the same setup as before. Thus, $\Lie_\zeta \no^a \=0$ whence $\kappao \=0$, and we have a family of cross-sections, $\uo ={\rm const}$, of $\scrip$ that are asymptotically shear-free in the distant past to which $\zeta^a$ is tangential and $\ello_a$ is orthogonal. Let us consider the portion $\scri^+_{[S_2,S]}$ of $\scrip$ bounded by a cross-section $S_2$ that belongs the preferred family and a generic cross-section $S$ to the future of $S_2$ which does not. In addition, $\zeta^a$ may not be tangential to $S$. Recall from section \ref{s2.1} that the charge $Q_\zeta [S]$ is well-defined for any $S$ and related to $Q_\zeta [S_2] = J_\zeta[S_2]$ by the general balance law (\ref{balance1}) at $\scrip$. Conditions satisfied by $\zeta^a$ and $\ello_a$ in our setup ensure that the flux on the right side of Eq.~(\ref{balance1}) assumes the simplified form (\ref{balance3}), whence we have
\ba J_\zeta [S_2] - Q_\zeta[S] \, &=& \, \f{1}{16\pi G} \int_{\scri^+_{[S_2,S]}}  \big[\Lie_\zeta\sigmao_{ab}\big]\, \No_{cd}\, \qo^{ac} \qo^{bd}\, \rmd^3 \Vo\, \nonumber\\
&=& \, \f{1}{16\pi G} \int_{\scri^+_{[S_2,S]}}  \big[\Lie_\zeta\sigma_{ab}\big]\, N_{cd}\, q^{ac} q^{bd}\, \rmd^3 V\,
\ea
where, in the second step we have expressed the right side using fields in the \AM conformal frame as in (\ref{eq:J_flux}). Combining this result with (\ref{balance4}) we have a balance law for any cross-section $S$ of $\scrip$:
\be \label{final} J_\zeta [\inot] - Q_\zeta[S] \, = \, \f{1}{16\pi G} \int_{\Delta\scrip} \big[\Lie_\zeta\sigma_{ab}\big]\, N_{cd}\, q^{ac} q^{bd}\, \rmd^3 V\, , \ee
where $\Delta\scrip$ now stands for the portion of $\scrip$ to the past of $S$. Note that $Q_\zeta [S]$ is just the $\zeta$-component of the angular momentum at the cross-section $S$; we use the stem letter $Q$ rather than $J$ simply because $J_\zeta [S]$ has been used in (\ref{scriJ1})  to denote the angular momentum associated with an ${\rm SO(3)}$ generator that is tangential to $S$. Since $\zeta$ is not tangential to $S$, we have to use the full expression (\ref{scriQ}) of the charge $Q_\zeta [S]$.}

2. In this calculation we focused on the ${\rm SO(3)}$ subgroup of $\inotP$ that is tailored to the asymptotically past rest-frame (i.e., that leaves invariant the $\admP$ and/or the past limit of $\bsP$) because angular momentum 3-vector $\vec{J}$ generally refers to this rotation group. But the analysis goes through for any {\rm $SO(3)$} subgroup of $\inotP$. Since we can express any generator $\xi^a$ of $\inotP$ as a linear combination of translations and generators of arbitrary ${\rm SO(3)}$ subgroups, and since the balance law also holds for the 4-momentum in \AM space-times 
\footnote{The argument in \cite{aaam-prl} for the 4-momentum balance law is similar to that in the main text in that one first extended structures from $\scrip\cup\inot$ to the space-time interior, considered the balance laws satisfied on cylinders $C$ and then took the limit $\omega\to 0$. However, there was an implicit interchange of the limit and the integral over $C$. To justify it one needs \AM boundary conditions that ensure that structures related to $\scrip$ and those related to $\inot$ are glued properly.}
the balance law holds for the full Lie algebra of $\inotP$. 

3. It is instructive to compare the discussion of this paper with that of \cite{aams-jmp} which established the relation between $\scriJ[S]$ and $\inotJ$ in stationary AEFANSI space-times. In that case two major simplifications occur: (i) there is a  4-parameter family of preferred cross-sections whose shear vanishes on \emph{entire} $\scrip$, and, (ii) Eq. (\ref{final}) reduces to  $J_\zeta [\inot] - J_\zeta[S] =0$ since $N_{ab} \=0 $ on all of $\scrip$. To establish this equality, one first chose the families of $\Omega = {\rm const.}$ cylinders $C$  as in Fig.~\ref{fig:foliation}, but such that the stationary Killing field $\hat{t}^a$ is tangential to them. One then extended the charge $J_\zeta [S]$ from $\scrip$ to the space-time interior using just the Weyl tensor, an extension $\tilde\zeta$ of the BMS rotation  $\zeta^a$, and of a null vector field $\hat\ell^a$ transverse to the cross-sections $\hat{S}$, both transported in a manner that respects the Killing symmetry. Then, the flux relating the charge integrals on 2-spheres $\hat{S_2}$ and $\hat{S}_0$ vanish already on cylinders $C$ and so there was no subtlety about the interchange of the limit $\Omega \to 0$ and integral of the flux over $C$. Thus, in stationary space-times one does 
not need the delicate \AM boundary conditions; the result holds in AEFANSI space-times. In non-stationary space-times, however, we do not have this luxury!

4. The balance law (\ref{final}) holds only because $\tilde\zeta^a$ is an asymptotic symmetry respecting the \AM structure and therefore belongs to $\inotP$ which correctly glues its action at $\inot$ with that on $\scrip$. To bring out this point, let us construct a vector field ${\tilde\xi}{}^a$ in space-time that is in the Lie-algebra of the \Poincare group $\spiP$, as well as of the BMS group  $\BMS$ on $\scrip$, but \emph{not} in the Lie-algebra of $\bmsP$, and investigate what happens to this balance law. An example of such a vector field  ${\tilde\xi}{}^a$ is obtained by adding to a rotation generator ${\tilde\zeta}{}^a$ in $\inotP$ a supertranslation $\mathfrak{s}^a$ on $\scrip$ that is \emph{not} a BMS translation:
\be \label{xi} {\tilde\xi}{}^a\, =\, \tilde\zeta^a + (1 - e^{- \f{\tilde\omega^4}{\Omega}})\, \tilde{\mathfrak{s}}^a \, \equiv\, \tilde\zeta^a \,+\, \mathfrak{f}\,\, \tilde{\mathfrak{s}}^a \,  \ee
where $\tilde{\mathfrak{s}}^a$ is a smooth extension of $\mathfrak{s}^a$ to the space-time interior (that is $C^{>0}$ at $\inot$ and smooth elsewhere), and  $\tilde\omega = \Omega/\Omegao$ relates the \AM conformal completion with the divergence free one. (Thus $\tilde\omega$ vanishes at $\inot$, is $C^{>0}$ there and smooth elsewhere.) Since  $\tilde\omega^2 \sim  \Omega$ in the space-like approach to $\inot$, it follows that ${\mathfrak{f}} \to 0$ exponentially in the limit to $\inot$ in space-like directions. On the other hand, since $\tilde\omega$ is strictly positive on $\scrip$, ${\mathfrak{f}}$ equals $1$ on all of $\scrip$. Therefore, the limit of ${\mathfrak{f}}$ to $\inot$ along any null direction is also $1$. Thus, the limit of ${\mathfrak{f}}$ to $\inot$ has a finite discontinuity along any of our families of `boosted' space-like curves $c_{\v}(\lambda)$ that converges to a curve $c_{\n}$ (with a null tangent vector $\n^a$ at $\inot$) as $\lambda \to 1$. As a consequence, the difference ${\tilde\xi}{}^a - {\tilde\zeta}{}^a$  goes to zero exponentially as one approaches $\inot$ along space-like directions, while we have ${\tilde\xi}{}^a - {\tilde\zeta}{}^a \,\=\, S^a$ everywhere on $\scrip$. Thus, as desired, although ${\tilde\xi}{}^a$ is an  asymptotic rotation in both $\spiP$ and $\BMS$, it  does not belong to $\bmsP$ because it differs from the $\tilde\zeta^a \in \bmsP$ by a \emph{supertranslation}. The difference goes to zero as one approaches $\inot$ along $\scrip$ simply because supertranslations $\mathfrak{s}^a$ goes to zero in this limit in any \AM conformal frame. Consequently,  although ${\mathfrak{f}}$ has a discontinuity, ${\tilde\xi}{}^a$ is continuous at $\inot$; it vanishes there along all directions. Nonetheless, as we show below, the discontinuity of $\mathfrak{f}$ makes a key difference.

Since this ${\tilde\xi}{}^a$ is a perfectly well-defined symmetry  in $\spiP$,  it defines an angular momentum charge $J_{\hat\xi} [\inot]$ at $\inot$ via (\ref{Jinot1}) or (\ref{Jinot2}) \emph{which equals} $J_{\hat\zeta} [\inot]$. On the other hand, it is also a well-defined symmetry in $\BMS$, with ${\tilde\xi}{}^a - {\tilde\zeta}{}^a = \mathfrak{s}^a$ on $\scrip$. Therefore, the definition (\ref{scriQ}) of BMS charges implies that  
\be \label{past-limit2} J_{\tilde\xi} [S]  - J_{\tilde{\zeta}} [S] =  Q_{\mathfrak{s}} [S]\, , \ee
where  $ Q_{\mathfrak{s}} [S]$ is the supermomentum charge associated with $S^a$ at the cross-section $S$. Consequently, in the infinite past limit in which the cross-section $S$ is sent to $\inot$, we have
\be \lim_{S \to \inot} J_{\tilde\xi} [S]  = J_{\tilde\zeta} [\inot] + \lim_{S \to \inot} Q_{\mathfrak{s}}[S]\, . \ee
Thus, the past limit of the angular momentum charge defined by ${\tilde\xi}{}^a$ via (\ref{scriQ}) does not in general equal the angular momentum charge it defines at $\inot$ via (\ref{Jinot1}) or (\ref{Jinot2}). Put differently, the equality between the past limit of $J_{\tilde\zeta}[S]$ and the angular momentum $J_{\tilde\zeta}$ defined at $\inot$ that we have for generators ${\tilde\zeta}{}^a$ of $\inotP$ falls to hold for ${\tilde\xi}{}^a$. 
This is because the diffeomorphism it generates fails to belong to $\inotP$: While ${\tilde\xi}{}^a$ belongs to Lie algebras of both $\spiP$ and $\BMS$, its limits along space-like directions are not sufficiently well-matched with those along null directions for its action to preserve the \AM structure.%
\footnote{Exception occurs if the past limit of the supermomentum $Q_{\mathfrak{s}}$ happens to vanish. This does happen for all supertranslations (that are not translations) if the past limit of the Newman-Penrose Weyl component $\Psi_2^\circ$ happens to be spherically symmetric in a Bondi conformal frame. In that `accidental case' the past limit of the angular momentum $J_{\tilde\xi} [S]$ does agree with angular momentum $J_{\tilde\xi}[\inot]$. This spherical symmetry may well be realized in binary coalescences but not in scattering situations. Even in cases where $\Psi_2^\circ$ happens to be spherically symmetric, conceptually it is not meaningful to compare $J_{\tilde\zeta} [\inot]$ with $J_{\tilde\xi} [S]$ via a balance law because they refer to different rotation groups. In terms of Newtonian mechanics, it would be like comparing $J_z$ of a particle in the distant past with, say, $J_y$ at a time $t$  to obtain a balance law in the accidental case when their values in the distant past happen to be the same.}

We can trace back our arguments to see why the equality (\ref{past-limit1}) fails for ${\tilde\xi}{}^a$: It is precisely because the discontinuity in ${\mathfrak{f}}$ prevents us to interchange the integration over $C$ and the limit $\Omega\to 0$. The expression of the linkage flux on cylinders $C$ involves \emph{derivatives} of the generator ${\tilde\xi}{}^a$, and they act on the discontinuous $\mathfrak{f}$ in the expression of ${\tilde\xi}{}^a$. Therefore, if one considers any of our families $C_{\v}(\lambda)$ of curves that converge to $c_{\n}$ as $\lambda \to 1$, the limit to $\inot$ of $\mathcal{F}_\zeta\, \rmd^3V (\lambda)$ fails to remain continuous, and so we cannot conclude that it is bounded in the neighborhood $\mathcal{N}$---and in fact it isn't. Hence, we cannot interchange the limit $\Omega\to 0$ and the integral over $C$. This example brings out the fact that even when ${\tilde\xi}{}^a$ preserves the asymptotic conditions in the two regimes separately --i.e., preserves the AEFANSI structure-- the balance law (\ref{final}) will generically fail unless it preserves the \AM boundary conditions and thus belongs to $\inotP$.

5. Recently, there has been renewed interest in the issue of angular momentum balance law especially in the context of scattering systems (see, e.g.,\cite{Damour:2020tta,Jakobsen:2021smu,Mougiakakos:2021ckm,Veneziano_2022,Manohar:2022dea,DiVecchia:2022owy,DiVecchia:2022piu}). Much of that discussion pertains to the post-Minkowskian approximation which has a \Poincare group that the angular momentum refers to. In \AM space-times that group is extended to the full theory by $\inotP$. Indeed, as Eq. (\ref{Jinot2}) shows, our angular momentum $\inotJ$ is the same as that constructed from initial data in these treatments. Therefore, our results could shed light on some issues of recent interest. First, as we saw, in generic scattering situations,  there is an exact angular momentum balance law (\ref{final}) \emph{provided} one chooses the rotation generators in the Lie algebra of $\inotP$. In particular, this implies that on $\scrip$, the generators must be tangential to a family of cross-sections of $\scrip$ \emph{that are asymptotically shear-free} in the distant past. Otherwise, the rotation generator would not belong to the Lie algebra of $\bmsP$ and one would be comparing angular momentum at spatial infinity using one ${\rm SO(3)}$ group, and that $\scrip$ using one that is supertranslated, leading to an `apples with oranges' type comparison. Second, there's some discussion about whether one can have angular momentum flux at $\scrip$ in absence of energy flux. Now if there is no energy flux across $\Delta\scrip$, then $N_{ab} =0$ there, whence it follows from~\eqref{final} that in the exact theory that the angular momentum flux associated with ${\tilde\zeta}{}^a$ must also vanish. Again this result holds because one uses the ${\rm SO(3)}$ group at $\scrip$ that is matched to that at $\inot$. For a generator of the type ${\tilde\xi}{}^a$ discussed in Remark 4 above, this matching fails and a spurious angular momentum flux may well appear if the past limit of supermomentum is non-zero. These are two illustrations of ways in which our results can clarify some of the issues that are currently debated in the context of scattering, using expansions in powers of Newton's constant.

6. For simplicity, we restricted our detailed discussion to the case when the physical metric $\hat{g}_{ab}$ satisfies source-free Einstein's equations in a neighborhood of $\scrip\cup \inot$. But our results go through also in the case when this condition is weakened to allow sources in this neighborhood, provided stress energy $\hat{T}_{ab}$ has the standard fall-off: $\Omega^{-2}\,\hat{T}_{ab}$ admits a $C^2$ limit to $\scrip$ and $\hat{T}_{ab}$ admits a regular direction dependent limit at $\inot$. As is well-known, the expressions of the BMS charges (\ref{scriQ}) and the $\inot$ charges (\ref{Jinot1}) remain unchanged. However, in the balance law (\ref{final}) the flux expression now has an additional matter contribution because, while only the Weyl part of the asymptotic curvature contributes to the charge aspect, to calculate the flux one needs to take its exterior derivative and simplify it using Bianchi identity on the Riemann tensor. This brings the Ricci part of the curvature, leading to an additional term representing the angular momentum carried by matter.
 
\section{Discussion}
\label{s4}

Let us begin with a summary, emphasizing conceptual issues. The BMS charges $Q_{\zeta}[S]$ and fluxes $F_{\zeta}[\Delta\scrip]$ at null infinity refer only to the BMS generators $\zeta^a$ and fields defined on $\scrip$, without any reference to the Spi group $\Spi$ or the asymptotic behavior of fields near $\inot$.  Reciprocally, charges $Q_{\tilde\zeta}[\inot]$  refer only to the generators $\xi^a$ of $\Spi$ and asymptotic fields at $\inot$ without any reference to the BMS group $\BMS$ or fields on $\scrip$. Therefore, in a space-time that satisfies asymptotic flatness conditions in the two regimes \emph{separately}, there is no relation between these sets of observables. To relate them one needs to restrict oneself to space-times in which, in addition, the structure at $\scrip$ and at $\inot$ are appropriately glued. This is achieved in part in AEFANSI space-times \cite{aarh,aa-ein} in which $\scrip$ arises as the future null cone of $\inot$. The BMS group $\BMS$ and the Spi group $\Spi$ each admits a unique translation subgroup and the AEFANSI structure suffices to identify each of these 4-dimensional Lie algebras with the tangent space $T_{\inot}$. This identification enables one to ask for the relation between the Bondi-Sachs and ADM 4-momenta $\bsP [S]$ and $\admP [\inot]$ as in \cite{aaam-prl}. 

For angular momentum, the situation is much more subtle because each of the symmetry groups $\BMS$ and $\Spi$ admits an \emph{infinite} number of Lorentz subgroups, related by supertranslations. Therefore, the minimal gluing provided by the AEFANSI boundary conditions does not suffice. In the companion paper \cite{aank1} we supplemented them by demanding certain continuity on that limits to $\inot$ along space-like and null directions to arrive at a definition of \emph{Asymptotically Minkowski} (\AM) space-times. The additional requirements are well-motivated from a geometric standpoint and rather weak. One does \emph{not} simply extend the $C^{>0}$ differentiability of the metric in an obvious manner. For example, the limits $\mathbf{E}_{ab}$ at $\inot$ of the electric part of the asymptotic Weyl curvature are guaranteed to exist along space-like directions already in AEFANSI space-times. Therefore, one might have expected that the continuity conditions of \AM space-times would imply that as one lets space-like directions approach a null direction (via a continuous family of increasing boosts), the limits would tend to a  finite value. This is not the case; generically they diverge\, (but the ADM 4-momentum constructed from a 2-sphere integral of $\mathbf{E}_{ab}$ with appropriate contractions remains well-defined in the limit). Similarly, if one passes from an \AM conformal completion to a more familiar divergence-free conformal completion (that is especially convenient if one is interested only in null infinity), then the Newman-Penrose component $\Psi_1^\circ$ of the Weyl curvature that features in the integrand of angular momentum at $\scrip$ is allowed to diverge in the distant past. (Again, the divergence is such that the angular momentum charge integral itself has a well-defined limit as one approaches $\inot$ along $\scrip$.) Thus, in particular one can use \AM space-times in scattering situations where the past limit of $\Psi_1^\circ$ generically diverges.  

As discussed in \cite{aank1}, while the boundary conditions satisfied by \AM space-times are rather tame in the sense that they are met in a wide class of space-times, they have interesting consequences. First, the news tensor (and radiative curvature components) automatically satisfy certain fall-off conditions as one approaches $\inot$ along $\scrip$ that have been generally imposed by hand to ensure that radiated angular momentum and energy is finite. The Second consequence is more surprising: they imply that the asymptotic symmetry group of these space-times is a \Poincare group $\inotP$ where the label $\inot$ is a reminder that it is selected by the gluing conditions at $\inot$. Since the primary interest of this paper lies in angular momentum, we focused on the generators ${\tilde\zeta}{}^a$ of ${\rm SO(3)}$ subgroups of $\inotP$.  Associated with any generator ${\tilde\zeta}{}^a$ of $\inot$, there is an angular momentum charge $J_{\tilde\zeta} [S]$ constructed from fields defined locally at a cross-section $S$ of $\scrip$, and also a charge $J_{\tilde\zeta}[\inot]$ constructed from limits of fields to $\inot$. Although in each regime the charge is defined without the knowledge of asymptotic flatness in the other, there is a balance law (\ref{final}): $J_{\tilde\zeta}[\inot]$ equals the sum of $J_{\tilde\zeta}[S]$ and the flux of angular momentum radiated across the portion $\Delta\scrip$ of $\scrip$ to the past of $S$. This relation holds only because the relevant structures that arise from approach to $\inot$ along space-like and null directions are glued appropriately at $\inot$.  To emphasize this point, in section \ref{s3.3.4} we considered a vector field ${\tilde\xi}{}^a$ that is a rotation generator separately in the two regimes but does not belong to the \Poincare group $\inotP$: It preserves the AEFANSI structure --i.e., asymptotic flatness in each regime \emph{separately}-- but not the \AM structure that glues the two appropriately. In that case, the balance law does not hold: in particular, the past limit of $J_{\tilde\xi} [S]$ does not equal $J_{\tilde\xi}[\inot]$ in generic situations. 

In this paper, we focused only on $\scrip$ and $\inot$. But the discussion shows that all our considerations apply also to $\scrim$. If we include $\scrim$, then one finds that there is a single symmetry group $\inotP$ that acts rigidly on $\scrim,\, \inot$ and $\scrip$. That is, the symmetry generator ${\tilde\zeta}{}^a$ belong to the sub-Lie algebra of $\spiP$ of the Spi group $\Spi$ at spatial infinity, and also to the sub-Lie algebras of the \Poincare subgroups $\bmsP$ of $\BMS$ on $\scrip$ as well as $\scrim$ (selected by the family of asymptotically shear-free cross-sections in each case). The infinitesimal transformations it generates unambiguously glue a given rotation on $\scrim$ to a specific rotation on $\scrip$ through $\inot$, without any supertranslation freedom. For such a rotation generator ${\tilde\zeta}{}^a$ the future limit  of the angular momentum $J_{\tilde\zeta} [\bar{S}]$ evaluated at a cross-section $\bar{S}$ of $\scrim$ also equals $J_{\tilde\zeta} [\inot]$, whence the future limit on $\scrim$ agrees with the past limit along $\scrip$. Since this holds for all rotation subgroups of $\inotP$, and since the same equality holds for the 4-momentum \cite{aaam-prl}, we conclude that the equality between future limits on $\scrim$ and past limits on $\scrip$ holds for all \Poincare charges associated with $\inotP$. 

However, our primary goal was to relate the Coulombic and radiative aspects of the gravitational field, i.e., structures at $\scrip$ with those at $\inot$ by examining how $\scrip$ is glued to $\inot$. Therefore, our focus has been on the relation between space-like and null approach to $\inot$ --relation between the $\scri^\pm$ is a byproduct. This is in contrast to some of the recent literature, where the primary goal has been to relate structures at $\scrim$ with those at $\scrip$ via an $S$-matrix, and $\inot$ serves only as an intermediate tool (see, in particular \cite{as-book,Prabhu:2019fsp,kpis,cgw,mohamed2023bmssupertranslation}). Much of that work is motivated by perturbative scattering theory. It has been known for quite some time that the enlargement of the \Poincare group in Minkowski space to the BMS group by supertranslations is tied to the infrared properties of the radiative aspects of non-perturbative quantum gravity \cite{ashtekar1987asymptotic,aa-yau,Ashtekar_2018}. Through Weinberg's soft theorems the recent work has made the relation much more detailed within perturbative quantum gravity \cite{as-book}. In this analysis, the BMS supermomenta and possible conservation laws from $\scrim$ to $\scrip$ for them play a key role. In \cite{Prabhu:2019fsp} a sufficient set of supplementary conditions was added to the definition of AEFANSI space-times to ensure that the conservation law holds for BMS supermomentum and subsequently, in \cite{kpis}, the result was extended to all BMS charges in classical general relativity. By contrast, the conservation law in the present \AM framework refers only to generators of $\inotP$; BMS supermomenta are not included. Could they be included by just taking the investigation further? As we now explain, answer seems to be in the negative for \AM space-times.

This may seem surprising at first since the notion of \AM space-times was also introduced by strengthening the AEFANSI asymptotic conditions. However, one can trace-back the reason to the differences in the way the conditions were strengthened. Since investigations in \cite{Prabhu:2019fsp,kpis} were primarily driven by conservation laws, it was natural to introduce a hierarchy of conditions that suffice to ensure that they hold. In particular, it was assumed that the Weyl curvature term in the \emph{integrand} of the BMS supermomentum charge admits (direction dependent) limits both along space-like and null directions and, furthermore, as the space-like directions are boosted to a null direction continuously, the limits along space-like directions approach the one along the null direction continuously (Eqs.~(2.41) and~(4.1) in \cite{Prabhu:2019fsp}.) In the \AM  framework, on the other hand, the AEFANSI boundary conditions were strengthened by first going back to the drawing board, so to say, whence the gluing procedure was guided primarily by geometric structures near $\inot$. As mentioned above, the motivation was to better understand the relation between radiative and Coulombic aspects that is encoded in the space-time geometry via Einstein's equations, rather than to arrive at $S$-matrix conservation laws. As a result, it turns out that the \AM boundary conditions do not imply the condition on the Weyl curvature term assumed in \cite{Prabhu:2019fsp,kpis}. In fact, in generic \AM space-times this condition is violated. Therefore, in striking contrast to the \Poincare charges associated with $\inotP$, there is no assurance that supermomentum would be conserved from $\scrip$ to $\scrim$.

Thanks to a recent result, one can make a more specific statement. Our results, as well as those of other investigations of this topic \cite{as-book,Prabhu:2019fsp,kpis,cgw}, do not address the issue of \emph{existence} of solutions to full non-linear Einstein's equations satisfying the assumed boundary conditions. Recently, this was remedied in part by using prior results of Ref.~\cite{Huang_2010} on solutions to the constraint equations with prescribed asymptotics. Suppose we are given an asymptotically flat solution with all the properties needed to examine conservation laws. Then Ref.~\cite{mohamed2023bmssupertranslation} provides explicit conditions that the Cauchy data on a 3-surface $\Sigma_0$ in that space-time must satisfy for the supermomentum conservation to hold: the angular dependence of the $1/r$-part of the asymptotic 3-metric is severely restricted; it has to satisfy an \emph{infinite} number of constraints! (See Proposition 2 and Theorem 1 in \cite{mohamed2023bmssupertranslation}). As a result, in the limit to $\inot$ along $\scripm$, half of the supermomentum charges on $\scripm$ are constrained to vanish! There is no obvious reason why this infinite set of constraints would be satisfied in all physically interesting situations, particularly those involving scattering in full classical general relativity.%
\footnote{Thus, there appears to be a healthy tension between perturbative results involving Weinberg's soft theorems and the exact, non-perturbative result of \cite{mohamed2023bmssupertranslation} in the classical theory. This should be a fertile area for further investigations given that, already in the classical theory, the situation vis a vis supermomentum conservation is quite different in the full general relativity and its linearized approximation \cite{Magdy_Ali_Mohamed_2022}. As a side remark, note that the result quoted above is a non-trivial generalization of the fact that if the initial data is asymptotically Schwarzschildean (as, e.g., in the Christodoulou-Klainnerman analysis of non-linear stability of Minkowski space), then the supermomentum conservation law holds rather trivially because the limit to $\inot$ along $\scripm$ of all `pure' supermomenta vanish.}  

Returning to \AM space-times, let us consider a Cauchy surface $\Sigma_0$ passing through $\inot$ where it is $C^{>1}$. If one examines the initial data induced by the physical metric $\hat{g}_{ab}$ on $\Sigma_0 \setminus \inot$, the angular dependence of the $1/r$-part of the 3-metric does not have to satisfy the constraints required for the supermomentum balance laws. Thus, \AM space-times appear to be general enough to accommodate generic scattering situations. (This expectation is also supported by the fact that in these space-times, the limit to $\inot$ of the Newman-Penrose curvature component $\Psi_1^\circ$ can diverge in the asymptotic past.) This is why in generic \AM space-times, although one does have a conservation law from $\scrim$ to $\scrip$ for the \Poincare momenta associated with $\inotP$, one does not expect one for supermomenta.

Let us conclude with a general observation. In the early days of research on gravitational radiation, it came as a major surprise that, even though the physical metric does approach a Minkowski metric near null infinity, the asymptotic symmetry group is not the \Poincare group but the BMS group. Now, physically it is clear that space-times representing isolated gravitating system should be asymptotically flat both at null and spatial infinity, not just separately, but in a way that unifies the two regimes seamlessly. \AM spacetimes provide a concrete example. But their asymptotic symmetry group takes us back to a \emph{\Poincare} group, $\inotP$! Although steps that led us to $\inotP$ seems logical and compelling, in view of the prominent role played by the BMS group in classical and quantum gravity, this reverting to the \Poincare group is likely to cause serious discomfort. Therefore, a clarification is warranted: our results do \emph{not} imply that one can just forgo the full BMS group $\BMS$ and work exclusively with $\inotP$. Irrespective of the issue of conservation laws between $\scrim$ and $\scrip$, supermomenta that $\BMS$ provides are physically interesting observables, and so are the associated soft charges and the gravitational memory. In the classical theory, this has been used to calculate the memory effect for numerical simulations~\cite{Mitman:2020bjf}, to calculate the posterior probability distributions for the gravitational memory in gravitational wave detections \cite{kkadl, Zhao:2021hmx}, and used to compare and contrast waveform models. On the quantum side, soft charges have been successfully used to highlight infrared issues in the asymptotic, non-perturbative quantization scheme \cite{ashtekar1987asymptotic}, and to relate them to the soft theorems of perturbative quantum gravity \cite{as-book}. Therefore, it is clear that the BMS group $\BMS$ will continue to be immensely useful. However, it is equally true that these successes of the BMS group do \emph{not} imply that ${\inotP}$ is physically unimportant or will not exist in physically interesting situations. Indeed, it is often present also in the analyses aimed at uncovering relation between fields on $\scrim$ and $\scrip$ that focus only on  the role of the BMS group (see, e.g., \cite{henneaux2,Prabhu:2019fsp,kpis,cgw,Capone:2022gme}). It's just that the fact that a preferred \Poincare subgroup exists when one joins $\scrip$ and $\scrim$ through $\inot$ is either not noticed or not emphasized. 

\section*{Acknowledgments}

We thank Juan Kroon and especially Kartik Prabhu for discussions and Gabriele Veneziano for motivating us to work on this issue.  This work was supported in part by the Eberly and Atherton research funds of Penn State and the Distinguished Visiting Research Chair program of the Perimeter Institute. N.K. is supported by the Natural Sciences and Engineering Council of Canada.

\bibliographystyle{JHEP}
\bibliography{ak2.bib}

\providecommand{\href}[2]{#2}\begingroup\raggedright\begin{thebibliography}{10}

\bibitem{aank1}
A.~Ashtekar and N.~Khera, \emph{{Unified Treatment of Null and Spatial Infinity III: Asymptotically Minkowski Space-times}},  \href{https://arxiv.org/abs/2311.14130}{{\ttfamily 2311.14130}}.

\bibitem{etnrp}
E.T.~Newman and R.~Penrose, \emph{Note on the bondi-metzner-sachs group}, {\emph{Journal of Mathematical Physics} {\bfseries 7} (1966) 863}.

\bibitem{aa-rad}
A.~Ashtekar, \emph{Radiative degrees of freedom of the gravitational field in exact general relativity}, {\emph{Journal of Mathematical Physics} {\bfseries 22} (1981) 2885}.

\bibitem{aarh}
A.~Ashtekar and R.O.~Hansen, \emph{{A unified treatment of null and spatial infinity in general relativity. I - Universal structure, asymptotic symmetries, and conserved quantities at spatial infinity}}, \href{https://doi.org/10.1063/1.523863}{\emph{J. Math. Phys.} {\bfseries 19} (1978) 1542}.

\bibitem{aa-ein}
A.~Ashtekar, \emph{Asymptotic structure of the gravitational field at spatial infinity}, {\emph{General Relativity and Gravitation II} {\bfseries 2} (1980) 37}.

\bibitem{bondi}
H.~Bondi, M.G.J.~Van~der Burg and A.~Metzner, \emph{Gravitational waves in general relativity, vii. waves from axi-symmetric isolated system}, {\emph{Proceedings of the Royal Society of London. Series A. Mathematical and Physical Sciences} {\bfseries 269} (1962) 21}.

\bibitem{sachs}
R.K.~Sachs, \emph{Gravitational waves in general relativity viii. waves in asymptotically flat space-time}, {\emph{Proceedings of the Royal Society of London. Series A. Mathematical and Physical Sciences} {\bfseries 270} (1962) 103}.

\bibitem{sachs2}
R.~Sachs, \emph{{Asymptotic symmetries in gravitational theory}}, \href{https://doi.org/10.1103/PhysRev.128.2851}{\emph{Phys. Rev.} {\bfseries 128} (1962) 2851}.

\bibitem{rp}
R.~Penrose, \emph{Zero rest-mass fields including gravitation: asymptotic behaviour}, {\emph{Proceedings of the Royal Society of London. Series A. Mathematical and physical sciences} {\bfseries 284} (1965) 159}.

\bibitem{adm}
R.~Arnowitt, S.~Deser and C.W.~Misner, \emph{Republication of: The dynamics of general relativity}, {\emph{General Relativity and Gravitation} {\bfseries 40} (2008) 1997}.

\bibitem{rg-jmp}
R.~Geroch, \emph{Structure of the gravitational field at spatial infinity}, {\emph{Journal of Mathematical Physics} {\bfseries 13} (1972) 956}.

\bibitem{Prabhu:2019fsp}
K.~Prabhu, \emph{Conservation of asymptotic charges from past to future null infinity: Supermomentum in general relativity}, {\emph{Journal of High Energy Physics} {\bfseries 2019} (2019) 1}.

\bibitem{kpis}
K.~Prabhu and I.~Shehzad, \emph{Conservation of asymptotic charges from past to future null infinity: Lorentz charges in general relativity}, {\emph{Journal of High Energy Physics} {\bfseries 2022} (2022) 1}.

\bibitem{am3+1}
A.~Ashtekar and A.~Magnon, \emph{From i° to the 3+ 1 description of spatial infinity}, {\emph{Journal of mathematical physics} {\bfseries 25} (1984) 2682}.

\bibitem{aaam-prl}
A.~Ashtekar and A.~Magnon-Ashtekar, \emph{Energy-momentum in general relativity}, {\emph{Physical Review Letters} {\bfseries 43} (1979) 181}.

\bibitem{aams-jmp}
A.~Ashtekar and M.~Streubel, \emph{On angular momentum of stationary gravitating systems}, {\emph{Journal of Mathematical Physics} {\bfseries 20} (1979) 1362}.

\bibitem{geroch1981linkages}
R.~Geroch and J.~Winicour, \emph{Linkages in general relativity}, {\emph{Journal of Mathematical Physics} {\bfseries 22} (1981) 803}.

\bibitem{winicour1968some}
J.~Winicour, \emph{Some total invariants of asymptotically flat space-times}, {\emph{Journal of Mathematical Physics} {\bfseries 9} (1968) 861}.

\bibitem{dray}
T.~Dray, \emph{Momentum flux at null infinity}, {\emph{Classical and Quantum Gravity} {\bfseries 2} (1985) L7}.

\bibitem{tdms}
T.~Dray and M.~Streubel, \emph{Angular momentum at null infinity}, {\emph{Classical and Quantum Gravity} {\bfseries 1} (1984) 15}.

\bibitem{adlkJ}
A.~Ashtekar, T.~De~Lorenzo and N.~Khera, \emph{Compact binary coalescences: The subtle issue of angular momentum}, {\emph{Physical Review D} {\bfseries 101} (2020) 044005}.

\bibitem{Compere:2019gft}
G.~Comp\`ere, R.~Oliveri and A.~Seraj, \emph{{The Poincar\'e and BMS flux-balance laws with application to binary systems}}, \href{https://doi.org/10.1007/JHEP10(2020)116}{\emph{JHEP} {\bfseries 10} (2020) 116} [\href{https://arxiv.org/abs/1912.03164}{{\ttfamily 1912.03164}}].

\bibitem{compere2021classical}
G.~Comp{\`e}re and D.A.~Nichols, \emph{Classical and quantized general-relativistic angular momentum}, {\emph{arXiv preprint arXiv:2103.17103} (2021) }.

\bibitem{Chen:2022fbu}
P.-N.~Chen, D.E.~Paraizo, R.M.~Wald, M.-T.~Wang, Y.-K.~Wang and S.-T.~Yau, \emph{{Cross-section continuity of definitions of angular momentum}}, \href{https://doi.org/10.1088/1361-6382/acaa82}{\emph{Class. Quant. Grav.} {\bfseries 40} (2023) 025007} [\href{https://arxiv.org/abs/2207.04590}{{\ttfamily 2207.04590}}].

\bibitem{aams}
A.~Ashtekar and M.~Streubel, \emph{Symplectic geometry of radiative modes and conserved quantities at null infinity}, {\emph{Proceedings of the Royal Society of London. A. Mathematical and Physical Sciences} {\bfseries 376} (1981) 585}.

\bibitem{chen2021supertranslation}
P.-N.~Chen, M.-T.~Wang, Y.-K.~Wang and S.-T.~Yau, \emph{Supertranslation invariance of angular momentum}, {\emph{arXiv preprint arXiv:2102.03235} (2021) }.

\bibitem{10.1063/1.525283}
A.~Ashtekar and J.~Winicour, \emph{Linkages and hamiltonians at null infinity}, {\emph{Journal of Mathematical Physics} {\bfseries 23} (1982) 2410}.

\bibitem{10.1063/1.531497}
R.~Beig and P.T.~Chruściel, \emph{{Killing vectors in asymptotically flat space–times. I. Asymptotically translational Killing vectors and the rigid positive energy theorem}}, \href{https://doi.org/10.1063/1.531497}{\emph{Journal of Mathematical Physics} {\bfseries 37} (1996) 1939} [\href{https://arxiv.org/abs/https://pubs.aip.org/aip/jmp/article-pdf/37/4/1939/8166635/1939\_1\_online.pdf}{{\ttfamily https://pubs.aip.org/aip/jmp/article-pdf/37/4/1939/8166635/1939\_1\_online.pdf}}].

\bibitem{10.4310/jdg/1669998184}
S.~Hirsch, D.~Kazaras and M.~Khuri, \emph{{Spacetime harmonic functions and the mass of 3-dimensional asymptotically flat initial data for the Einstein equations}}, \href{https://doi.org/10.4310/jdg/1669998184}{\emph{Journal of Differential Geometry} {\bfseries 122} (2022) 223 }.

\bibitem{Damour:2020tta}
T.~Damour, \emph{{Radiative contribution to classical gravitational scattering at the third order in $G$}}, \href{https://doi.org/10.1103/PhysRevD.102.124008}{\emph{Phys. Rev. D} {\bfseries 102} (2020) 124008} [\href{https://arxiv.org/abs/2010.01641}{{\ttfamily 2010.01641}}].

\bibitem{Jakobsen:2021smu}
G.U.~Jakobsen, G.~Mogull, J.~Plefka and J.~Steinhoff, \emph{{Classical Gravitational Bremsstrahlung from a Worldline Quantum Field Theory}}, \href{https://doi.org/10.1103/PhysRevLett.126.201103}{\emph{Phys. Rev. Lett.} {\bfseries 126} (2021) 201103} [\href{https://arxiv.org/abs/2101.12688}{{\ttfamily 2101.12688}}].

\bibitem{Mougiakakos:2021ckm}
S.~Mougiakakos, M.M.~Riva and F.~Vernizzi, \emph{{Gravitational Bremsstrahlung in the post-Minkowskian effective field theory}}, \href{https://doi.org/10.1103/PhysRevD.104.024041}{\emph{Phys. Rev. D} {\bfseries 104} (2021) 024041} [\href{https://arxiv.org/abs/2102.08339}{{\ttfamily 2102.08339}}].

\bibitem{Veneziano_2022}
G.~Veneziano and G.A.~Vilkovisky, \emph{Angular momentum loss in gravitational scattering, radiation reaction, and the bondi gauge ambiguity}, \href{https://doi.org/10.1016/j.physletb.2022.137419}{\emph{Physics Letters B} {\bfseries 834} (2022) 137419}.

\bibitem{Manohar:2022dea}
A.V.~Manohar, A.K.~Ridgway and C.-H.~Shen, \emph{{Radiated Angular Momentum and Dissipative Effects in Classical Scattering}}, \href{https://doi.org/10.1103/PhysRevLett.129.121601}{\emph{Phys. Rev. Lett.} {\bfseries 129} (2022) 121601} [\href{https://arxiv.org/abs/2203.04283}{{\ttfamily 2203.04283}}].

\bibitem{DiVecchia:2022owy}
P.~Di~Vecchia, C.~Heissenberg and R.~Russo, \emph{{Angular momentum of zero-frequency gravitons}}, \href{https://doi.org/10.1007/JHEP08(2022)172}{\emph{JHEP} {\bfseries 08} (2022) 172} [\href{https://arxiv.org/abs/2203.11915}{{\ttfamily 2203.11915}}].

\bibitem{DiVecchia:2022piu}
P.~Di~Vecchia, C.~Heissenberg, R.~Russo and G.~Veneziano, \emph{{Classical gravitational observables from the Eikonal operator}}, \href{https://doi.org/10.1016/j.physletb.2023.138049}{\emph{Phys. Lett. B} {\bfseries 843} (2023) 138049} [\href{https://arxiv.org/abs/2210.12118}{{\ttfamily 2210.12118}}].

\bibitem{as-book}
A.~Strominger, \emph{Lectures on the infrared structure of gravity and gauge theory}, Princeton University Press (2018).

\bibitem{cgw}
G.~Comp{\`e}re, S.E.~Gralla and H.~Wei, \emph{An asymptotic framework for gravitational scattering}, {\emph{arXiv preprint arXiv:2303.17124} (2023) }.

\bibitem{mohamed2023bmssupertranslation}
M.M.A.~Mohamed, K.~Prabhu and J.A.V.~Kroon, \emph{{BMS-supertranslation charges at the critical sets of null infinity}},  \href{https://arxiv.org/abs/2311.07294}{{\ttfamily 2311.07294}}.

\bibitem{ashtekar1987asymptotic}
A.~Ashtekar, \emph{Asymptotic quantization: based on 1984 naples lectures}, {\emph{Bibliopolis, Naples} (1987) }.

\bibitem{aa-yau}
A.~Ashtekar, \emph{{Geometry and Physics of Null Infinity}}, {\emph{{Surveys in Differential Geometry, a Jubilee Volume on General Relativity and Mathematics celebrating 100 Years of General Relativity}} (2014) } [\href{https://arxiv.org/abs/1409.1800}{{\ttfamily 1409.1800}}].

\bibitem{Ashtekar_2018}
A.~Ashtekar, M.~Campiglia and A.~Laddha, \emph{Null infinity, the bms group and infrared issues}, \href{https://doi.org/10.1007/s10714-018-2464-3}{\emph{General Relativity and Gravitation} {\bfseries 50} (2018) }.

\bibitem{Huang_2010}
L.-H.~Huang, \emph{{Solutions of special asymptotics to the Einstein constraint equations}}, \href{https://doi.org/10.1088/0264-9381/27/24/245002}{\emph{Class. Quant. Grav.} {\bfseries 27} (2010) 245002} [\href{https://arxiv.org/abs/1002.1472}{{\ttfamily 1002.1472}}].

\bibitem{Magdy_Ali_Mohamed_2022}
M.M.A.~Mohamed and J.A.V.~Kroon, \emph{{Asymptotic charges for spin-1 and spin-2 fields at the critical sets of null infinity}}, \href{https://doi.org/10.1063/5.0081834}{\emph{J. Math. Phys.} {\bfseries 63} (2022) 052502} [\href{https://arxiv.org/abs/2112.03890}{{\ttfamily 2112.03890}}].

\bibitem{Mitman:2020bjf}
K.~Mitman et~al., \emph{{Adding gravitational memory to waveform catalogs using BMS balance laws}}, \href{https://doi.org/10.1103/PhysRevD.103.024031}{\emph{Phys. Rev. D} {\bfseries 103} (2021) 024031} [\href{https://arxiv.org/abs/2011.01309}{{\ttfamily 2011.01309}}].

\bibitem{kkadl}
N.~Khera, B.~Krishnan, A.~Ashtekar and T.~De~Lorenzo, \emph{Inferring the gravitational wave memory for binary coalescence events}, {\emph{Physical Review D} {\bfseries 103} (2021) 044012}.

\bibitem{Zhao:2021hmx}
Z.-C.~Zhao, X.~Liu, Z.~Cao and X.~He, \emph{{Gravitational wave memory of the binary black hole events in GWTC-2}}, \href{https://doi.org/10.1103/PhysRevD.104.064056}{\emph{Phys. Rev. D} {\bfseries 104} (2021) 064056} [\href{https://arxiv.org/abs/2111.13882}{{\ttfamily 2111.13882}}].

\bibitem{henneaux2}
M.~Henneaux and C.~Troessaert, \emph{Hamiltonian structure and asymptotic symmetries of the einstein-maxwell system at spatial infinity}, {\emph{Journal of High Energy Physics} {\bfseries 2018} (2018) 1}.

\bibitem{Capone:2022gme}
F.~Capone, K.~Nguyen and E.~Parisini, \emph{{Charge and antipodal matching across spatial infinity}}, \href{https://doi.org/10.21468/SciPostPhys.14.2.014}{\emph{SciPost Phys.} {\bfseries 14} (2023) 014} [\href{https://arxiv.org/abs/2204.06571}{{\ttfamily 2204.06571}}].

\end{thebibliography}\endgroup

\end{document}